\documentclass[a4paper,11pt]{article}
\pdfoutput=1 

\usepackage{jcappub}
\usepackage{graphicx}
\usepackage{hyperref}
\usepackage{aas_macros}
\usepackage{amsmath}
\usepackage{amssymb}
\usepackage{xcolor}
\usepackage{nccmath}

\newcommand{\GeV}{~\text{GeV}}

\newcommand{\Msol}{~M_{\odot}}

\newcommand{\vir}{\mathrm{vir}}
\newcommand{\pad}{\hspace{3mm}}

\def\beq{\begin{equation}\begin{aligned}}
\def\eeq{\end{aligned}\end{equation}}

\begin{document}

\title{\boldmath  Dark Black Holes in the Mass Gap}

\author[1,2]{Nicolas~Fernandez,}
\author[3,4,5]{Akshay~Ghalsasi,}
\author[3,4]{Stefano~Profumo,}
\author[3,6]{Lillian~Santos-Olmsted}
\author[3,4]{and Nolan~Smyth}

\affiliation[1]{Department of Physics, University of Illinois at Urbana-Champaign, Urbana, IL 61801, USA}
\affiliation[2]{ Illinois Center for Advanced Studies of the Universe, University of Illinois at Urbana-Champaign, Urbana, IL 61801, USA}
\affiliation[3]{Department of Physics, 1156 High St., University of California Santa Cruz, Santa Cruz, CA 95064, USA}
\affiliation[4]{Santa Cruz Institute for Particle Physics, 1156 High St., Santa Cruz, CA 95064, USA}
\affiliation[5]{%
Pittsburgh Particle Physics, Astrophysics, and Cosmology Center, Department of Physics and Astronomy,
University of Pittsburgh, Pittsburgh, USA
}%
\affiliation[6]{Department of Physics, Stanford University, Stanford, CA 94305, USA}

 \emailAdd{nicofer@illinois.edu}
\emailAdd{akg53@pitt.edu}
 \emailAdd{profumo@ucsc.edu}
 \emailAdd{nwsmyth@ucsc.edu}
 \emailAdd{lsantoso@ucsc.edu}

\abstract{In the standard picture of stellar evolution, pair-instability -- the energy loss in stellar cores due to electron-positron pair production -- is predicted to prevent the collapse of massive stars into black holes with mass in the range between approximately 50 and 130 solar masses -- a range known as the ``{\em black hole mass gap}''. LIGO and Virgo detection of black hole binary mergers containing one or both black holes with masses in this {\em mass gap} thus challenges the standard picture, possibly pointing to an unexpected merger history, unanticipated or poorly understood astrophysical mechanisms, or new physics. Here, we entertain the possibility that a ``dark sector'' exists, consisting of dark electrons, dark protons, and electromagnetic-like interactions, but no nuclear forces. Dark stars would inevitably form given such dark sector constituents, possibly collapsing into black holes with masses within the mass gap. We study in detail the cooling processes necessary for successful stellar collapse in the dark sector and show that for suitable choices of the particle masses, we indeed predict populating the mass gap with dark sector black holes. In particular, we numerically find that the heavier of the two dark sector massive particles cannot be lighter than, approximately, the visible sector proton for the resulting dark sector black holes to have masses within the mass gap. We discuss constraints on this scenario and how to test it with future, larger black hole merger statistics.}

\maketitle
\flushbottom
%%%%%%%%%%%%%%%%%%%%%%%%%%%%%%%%%%%%%%%%%%%%%%%%%%%%%
%%%%%%%%%%%%%%%%%%%%%%%%%%%%%%%%%%%%%%%%%%%%%%%%%%%%%
\section{Introduction}
\label{sec:intro}
%%%%%%%%%%%%%%%%%%%%%%%%%%%%%%%%%%%%%%%%%%%%%%%%%%%%%
%%%%%%%%%%%%%%%%%%%%%%%%%%%%%%%%%%%%%%%%%%%%%%%%%%%%%
The momentous discovery of gravitational waves \cite{PhysRevLett.116.061102} ushered a new era in astronomy and astrophysics with implications that could impact the fundamental picture of particles and their interactions. As the statistics of binary mergers detected via their gravitational radiation keep increasing \cite{LIGOScientific:2021psn}, population studies, especially of massive black holes, are becoming increasingly statistically meaningful. One of the most puzzling findings at present is the absence of the expected ``{\em mass gap}'' - the supposed absence of black holes in the mass range between roughly 50 and 130 solar masses \cite{Woosley_2021}. As explained below, the mass gap is predicted for black holes with a standard stellar collapse origin; while {\em per se} the absence of a mass gap might well indicate the breakdown of some of the assumptions leading to its prediction from stellar evolution, and/or an unexpected merger history of black hole populations at early times, it might also point to a non-stellar origin for the black holes populating the predicted gap. If that is the case, such black holes might be {\em primordial} -- not originating from stellar collapse, but rather from the collapse of early density fluctuations dense enough that both the size of the perturbation is larger than the Jeans scale and the particle horizon is larger than the gravitational radius \cite{Carr:2009jm}. Forming primordial black holes with masses in the ``mass gap'' is, however, very problematic, as the required large density fluctuations would need to arise at very late times (although see Refs~\cite{Ashoorioon:2019xqc, Ashoorioon:2022raz}). As such, the existence of ``primordial'' black holes in the mass gap might point to an entirely different formation pathway, for instance the one we entertain here (and that was proposed, in a different context, before, see e.g. \cite{Shandera:2018xkn}): the black holes in the mass gap might originate from the collapse of ``dark stars'' formed by the collapse of dark sector ``electrons'' and ``protons'' that are not supported by any dark nuclear interactions. Before exploring this admittedly rather unconventional scenario, we briefly review the theoretical underpinning for the existence of a mass gap in the standard stellar collapse picture.

The black hole mass gap (BHMG) is a result of the pair instability caused by the production of electron-positron pairs in stellar cores, which converts a portion of radiation pressure into massive particles, reducing the overall outward pressure support the star \cite{Rakavy_1967}. The resulting implosion is reversed by oxygen burning \cite{PhysRevLett.18.379}, leading to one of two possible outcomes: Stars with an initial mass $70 \Msol \lesssim M_{\rm{in}} \lesssim 140 \Msol$ undergo a phase known as “pulsational pair-instability supernova”, consisting of pulsations in which their cores contract, burn, expand, cool, and then contract again, ejecting mass in each pulsation. It is generally believed that these cores eventually collapse into black holes of mass $35 \Msol \lesssim M_{\rm{BH}} \lesssim 50 \Msol$ \cite{Woosley_2017}. For stars with an initial mass $140 \Msol \lesssim M_{\rm{in}} \lesssim 260 \Msol$, the explosion following the implosion is so violent that the entire star is disrupted and no remnant is left behind. This explosion is the “pair-instability supernova” and it accounts for the existence of the mass gap. Stars with initial masses $\gtrsim  260 \Msol$ have insufficient nuclear energy to counteract the implosion caused by the pair-instability. Consequently, their cores collapse into black holes of mass $\gtrsim 130 \Msol$ \cite{Heger_2002}. Given that the pair-instability supernova regime is bounded both from above and from below by the fact that stars form black holes outside this range, the generic expectation is {\em not to observe any stellar black holes of mass $50 \Msol \lesssim M_{\rm{BH}} \lesssim 130 \Msol$} \cite{Woosley_2021}.

Intriguingly, some of the merger events observed by LIGO \cite{LIGOScientific:2014pky} and Virgo \cite{VIRGO:2014yos} in 2019 have indicated the existence of black holes in the mass gap with strong statistical significance  \cite{LIGOScientific:2021usb}. The heavier of the original masses in event GW190521 is $98.4^{+33.6}_{-21.7} \ \Msol$ with the lighter mass of $57.2^{+27.1}_{-21.7} \ \Msol$; the merger masses for GW200220\textunderscore061928 are $87^{+40}_{-23} \ \Msol$ and $61^{+26}_{-25} \ \Msol$. Additionally, the primary black hole mass in the events GW190602\textunderscore175927, GW190706\textunderscore222641 and GW190929\textunderscore012149 are $71.8^{+18.1}_{-14.6} \ \Msol$, $74.0^{+20.1}_{-16.9} \ \Msol$ and  $66.3^{+21.6}_{-16.6} \ \Msol$ respectively \cite{LIGOScientific:2021usb,LIGOScientific:2021djp}. As alluded to above, these events suggest that there may be some unknown black hole formation mechanism not described by our current models of stellar evolution. Among credible astrophysical mechanisms that could fill the mass gap we mention here second-generation BHs emerging from previous merger of two BHs \cite{Gerosa:2017kvu, Fishbach:2017dwv, Rodriguez:2019huv, LIGOScientific:2020ufj}, accretion inside compact gaseous proto{-}clusters \cite{Roupas:2018cvb, Roupas:2019dgx}, super-Eddington accretion in isolated binaries \cite{vanSon:2020zbk}, or mergers
between an evolved star and a main-sequence companion
 \cite{Spera:2018wnw, DiCarlo:2019pmf}. 

The work in explaining black holes in the BHMG using beyond Standard Model physics has thus far primarily focused on additional cooling mechanisms in the star \cite{Croon:2020ehi,Croon:2020oga,Sakstein:2020axg,Ziegler:2020klg}, which prevent the pulsation pair instability supernova. In this paper we present an alternative. We consider the possibility that a fraction of dark matter consists of a dissipative component, specifically a ``dark atom" with a heavy particle (we refer to as the ``dark proton") with mass $m_{X}$, a lighter particle (``dark electron'') with mass $m_{c} < m_{X}$, and a dark fine-structure constant $\alpha_{D}$. Such a complex dark sector is well motivated by solutions to the hierarchy problem (e.g. Twin Higgs, $N$-Naturalness) \cite{Chacko:2005pe,Arkani-Hamed:2016rle}. However we will remain agnostic to the exact origin of the dark sector.

We consider the cooling and fragmentation of such a dark sector and show that this can lead to the formation of ``dark stars". We will use the term ``stars" to describe these dark gas clumps since they evolve according to the typical equations of stellar evolution. However, we assume that there is no nuclear physics in the dark sector and stress that our use of the word ``stars" should not be understood to imply the presence of nuclear reactions. We will investigate the parameter space in the $m_{X}, m_{c} ,\alpha_{D}$ plane that will form black holes in the BHMG.

The likelihood of our atomic dark sector forming black holes within the BHMG depends on the following considerations. As structure forms hierarchically, the atomic dark matter falls into virialized halos and shock heats to the virial temperature. If the virial temperature is above the ionization energy of atoms, the dark atoms will ionize and cool through Compton scattering, bremsstrahlung, and atomic transitions. If a significant amount of molecules form during this process, then molecular processes can efficiently cool the gas further. As the atomic dark sector cools, it fragments to smaller mass scales. This continues until the opacity reaches a point where it cannot cool anymore and the gas stops fragmenting. The smallest scale of fragmentation reached when the gas reaches thermal equilibrium (TE) is called the ``Jeans Mass'', $M_{J}$.

The Jeans mass of the star is, in turn, controlled by the cooling mechanism (atomic vs. molecular) that allows for the smallest Jeans mass at a given point in the parameter space. Finally, once the star becomes opacity limited, it reaches hydrostatic equilibrium and further evolution into either a black hole or a degeneracy supported star (``white dwarf'') is determined by its mass. Assuming the star is heavier than the Chandrasekhar mass, the star contracts by radiating away energy from its surface until the core temperature becomes comparable to the electron mass, at which point a heavy enough star will experience pair instability and undergo dynamic collapse into a BH. We also verify that our dark stars can form a black hole before $z = 2$ and that the stars will lose pressure support, i.e. the first adiabatic index, $\Gamma_{1}$, drops below $4/3$ (see Section \ref{sec:evolution} and Appendix \ref{sec:eos}).

We discuss the above considerations in detail in the remainder of this paper, which is structured as follows: In Section \ref{sec:adm} we  describe the paradigm of ``atomic dark matter" and define reasonable boundaries on the dark electron mass $m_{c}$, the dark proton mass $m_{X}$, and the dark fine-structure constant, $\alpha_{D}$. We also determine the parameter space where cooling via atomic cooling (including atomic transitions, Compton scattering and bremsstrahlung) is efficient. In Section \ref{sec:mj} we compute the Jeans mass of the dark stars assuming atomic and molecular cooling, and derive the parameter space that gives rise to black holes in the mass gap. In Section \ref{sec:evolution} we consider the evolution of dark stars, and describe the parameter space where the stars can cool efficiently within the age of the universe until dynamic collapse sets in. We also derive the adiabatic index of the core to find the mass of stars for a given parameter space that experiences core collapse. In Section \ref{sec:constraints} we discuss other constraints on the dark atom parameter space. We summarize our findings and discuss future directions in Section  \ref{sec:conclusions}.

%%%%%%%%%%%%%%%%%%%%%%%%%%%%%%%%%%%%%%%%%%%%%%%%%%%%%
%%%%%%%%%%%%%%%%%%%%%%%%%%%%%%%%%%%%%%%%%%%%%%%%%%%%%
\section{Atomic Dark Matter}
\label{sec:adm}
%%%%%%%%%%%%%%%%%%%%%%%%%%%%%%%%%%%%%%%%%%%%%%%%%%%%%
%%%%%%%%%%%%%%%%%%%%%%%%%%%%%%%%%%%%%%%%%%%%%%%%%%%%%
As discussed above, a fraction of dark matter can form atomic states consisting of a heavy ``proton" with mass $m_{X}$ and a light electron with mass $m_{c} \ll m_{X}$ and we denote the fraction of atomic dark matter (ADM) $\epsilon = \frac{\Omega_{ADM}}{\Omega_{DM}}$. Atomic dark matter and the associated phenomenology  has been explored extensively \cite{Goldberg:1986nk,Kaplan:2009de, Kaplan:2011yj, Cyr-Racine:2012tfp, Cline:2012is,Cline:2013pca, Fan:2013yva,Fan:2013tia,Fan:2013bea,Cyr-Racine:2013fsa,2014,Foot:2014osa,Foot:2014uba, Foot:2016wvj,Rosenberg:2017qia, Ghalsasi:2017jna,Chang:2018bgx,Shandera:2018xkn,Gresham:2018anj,Essig:2018pzq,Alvarez:2019nwt,Roux:2020wkp,Cyr-Racine:2021oal,Cline:2021itd,Chacko:2021vin,Ryan:2021dis,Gurian:2021qhk,Ryan:2021tgw,Howe:2021neq,Blinov:2021mdk,Bansal:2021dfh,Cruz:2022otv,Peled:2022byr,Roy:2023zar, Gemmell:2023trd}. Dark stars as well as dark compact objects such as dark white dwarfs and dark neutron stars have been considered in \cite{Kouvaris:2015rea, Giudice:2016zpa, Curtin:2019lhm, Hippert:2021fch, Gross:2021qgx, Ryan:2022hku, Gurian:2022nbx, Armstrong:2023cis}.

The upper bound on the fraction of atomic dark matter comes from considering the impact of the inevitable dark acoustic oscillations on the CMB as well as on the matter power spectrum \cite{Cyr-Racine:2012tfp,Cyr-Racine:2013fsa}. While the exact constraints depend on the ratio of the dark CMB temperature to the visible-sector CMB temperature, as well as on $m_{X}, m_{c}, \alpha_{D}$, all of the parameter space can be accommodated as long as the dark atoms make up at most $\approx 5\%$ of the total DM density. 
In Ref.~\cite{Ghalsasi:2017jna} galaxy and star formation from dark atoms was investigated and constraints on the parameter space were derived assuming that $5\%$ of the dark matter is atomic. Strong constraints from measurements of galaxy mass (from rotation curves), as well as dynamic heating of ultra-faint dwarfs from black holes formed from the dark sector rule out most of the interesting parameter space of atomic dark matter. However in this paper we assume that atomic dark matter can generically be much less than $5\%$ of the dark matter density, potentially avoiding the above-mentioned constraints. 

\begin{figure}
    \centering
    \includegraphics[width=0.495\columnwidth]{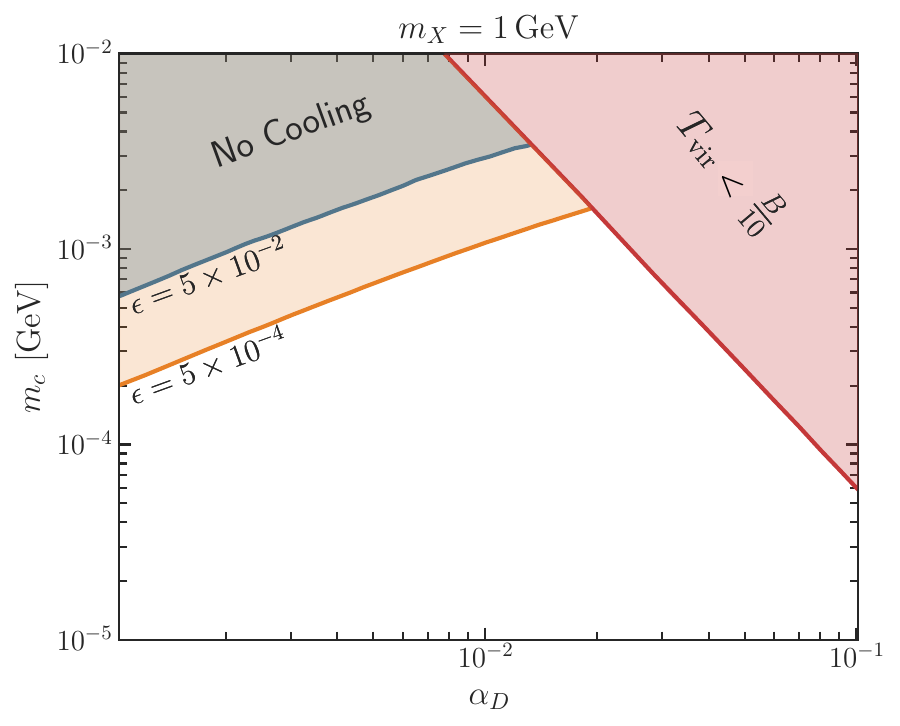}
    \includegraphics[width=0.495\columnwidth]{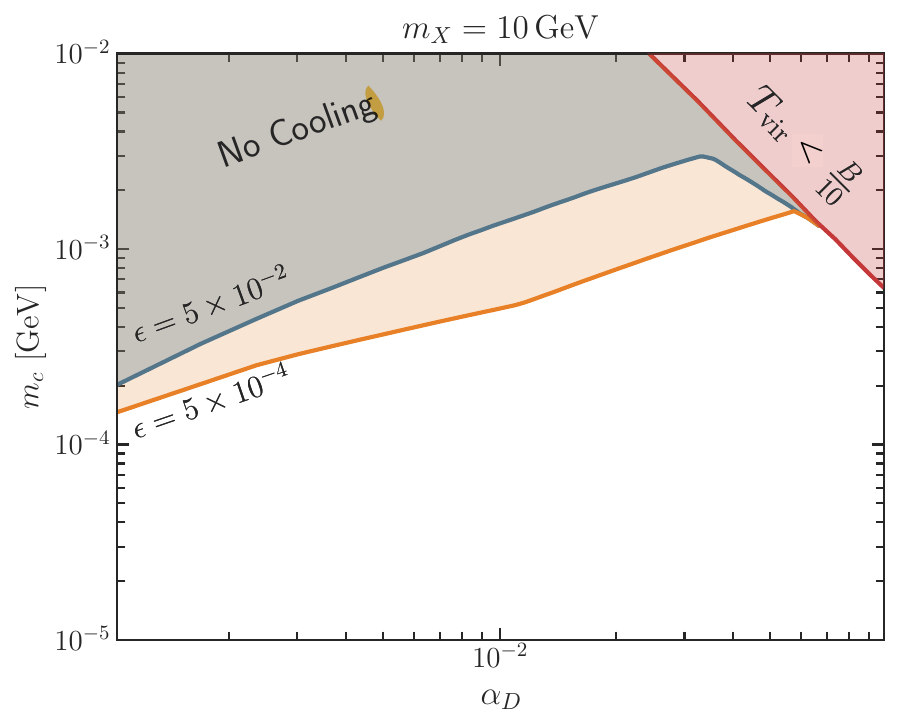}
    \caption{\textbf{Left:} Parameter space in which dark atoms can cool efficiently via atomic transitions, bremsstrahlung and Compton scattering for $m_{X} = 1\GeV$. \textbf{Right:} Parameter space in which dark atoms can cool efficiently via atomic transitions, bremsstrahlung and Compton scattering for $m_{X} = 10\GeV$.}
    \label{fig:tcool}
\end{figure}

Let us now discuss the parameter space $(m_{X},m_{c},\alpha_{D})$. The dark atoms fall into halos formed by dark matter, shock-heating to the virial temperature of the halo $T_{\vir}$. If the virial temperature is comparable or larger than the binding energy of the atom, $B = \alpha^{2}_{D} m_{c}/2$, then the dark hydrogen ionizes and cools via Compton scattering, bremsstrahlung, and atomic transitions. The cooling time has to be fast enough that the gas cools before the halo experiences further mergers which increase its virial temperature. Most of the parameter constraints we mention below depend on the ability of the gas to cool efficiently. We discuss the cooling and other relevant timescales in detail in Appendix \ref{sec:appcoolingtimes} and justify our parameter choice. Our parameter range choices are as follows:

\begin{itemize}
    \item{\textbf{$\epsilon \in [5 \times 10^{-4}, 5\times10^{-2}]$}. 
    Here the upper bound of $5\%$ comes from the discussion above (and in \cite{Ghalsasi:2017jna}) while the lower bound comes from the ability of the gas to equilibrate within the lifetime of the universe\footnote{Compton scattering does not depend on $\epsilon$ and hence can be always efficient in a large part of the parameter space, thus the constraint comes from the ability of gas to equilibrate. See Fig.~\ref{fig:tcool} and Appendix~\ref{sec:appcoolingtimes} for details.}.}
    
    \item{\textbf{$\alpha_{D} \in [10^{-3}, 10^{-1}]$}. 
    Here the lower bound on $\alpha_{D}$ comes from inefficient cooling. The upper bound comes from the fact that we can use standard perturbation theory results for our hydrogen atoms and molecules as well as keeping the binding energy low enough to allow the gas to ionize after virialization.}
    
    \item{$m_{c} \in [10^{-5} \GeV, 10^{-2} \GeV]$. 
    Here the upper bound stems from inefficient cooling and the need to keep the binding energy lower than the virialization temperature. The lower bound comes from the need to keep the electrons non-relativistic at high virial temperatures.}
    
    \item{$m_{X} \in [1 \GeV, 10 \GeV]$. Here the lower bound comes from requiring that the dark stars we want to form have masses in the mass gap $\mathcal{O}(100)\Msol$ which is below the Chandrasekhar mass, $M_{c} \approx 5.6 \Msol \left(\frac{\GeV}{m_{X}}\right)^{2}$ \footnote{The Chandrasekhar mass is $5.6 \Msol$ for pure hydrogen stars.} if $m_{X} \leq 0.1 \GeV$ . The upper bound comes from considering the thermal equilibration timescale between electrons and protons which can be prohibitively large for $m_{X} \gg 10 \GeV$.}
    
    \item{Finally we fix the ratio of dark CMB temperature to CMB temperature to be $r = \frac{T_{d,0}}{T_{0}} = 0.5$ \footnote{Note that $r \leq  0.5$ is required by current CMB measurement. None of our results depend on the actual value of $r$ except for a small part of the parameter space that cools by Compton cooling.}.}
    
\end{itemize}
\subsection{Atomic cooling}
In order for atomic dark matter to cool efficiently, the gas has to be partially ionized. Hence, the virial temperature has to satisfy $T_{\vir} \geq \frac{B_{X}}{10}$ \footnote{It is possible that for $T_{\vir} < B_{X}/10$ the gas can cool through molecular cooling from molecules formed during dark recombination. However molecular cooling is quadrupole ($\alpha^{2}_{D}$) suppressed  as well as $\left(m_{c}/m_{X}\right)^{5}$ suppressed, compared to atomic cooling. Hence we ignore this possibility.}. During hierarchical structure formation, the virial temperature of the halos grows with decreasing redshift. Cooling is more efficient at larger redshifts because of the correspondingly larger number density. Thus the cooling rate is maximal at the redshift $z$ when $T_{\vir} \approx \frac{B_{X}}{10}$.   In Appendix \ref{sec:appcoolingtimes} we provide a semi-analytical method to calculate the regions of our $\alpha_{D}, m_{c}$ parameter space that will cool at least $10 \%$ of atomic dark matter efficiently for a given $m_{X}$. Since the black holes observed by LIGO, Virgo and KAGRA have $z < 2$, we will require the gas to cool by $z = 2~(t = 3.3 {~\rm Gyr})$.

\subsection{Molecular Cooling}

Dark molecules, the bound states of dark atoms, can cool the gas to even lower temperatures. The ability for dark molecules to cool depends on the fraction of dark molecules that can form. However, it is not straightforward to predict the fraction of dark molecules that form in the universe \footnote{For e.g. at low redshifts hydrogen molecules form on dust grains in our universe.}. Even taking into account equilibrium processes, hydrogen molecule formation is an involved network of creation and destruction processes that are difficult to approximate analytically. Moreover $H^{-}$, which catalyzes the formation of $H_{2}$, has a low binding energy of $B_{X}/10$ and can be easily destroyed by radiation from the first dark stars. Important progress has been made recently in calculating the formation of dark molecular hydrogen quantitatively \cite{Ryan:2021dis,Gurian:2021qhk,Ryan:2021tgw}. Such a calculation needs to be embedded into a merger tree or N-body simulation in order to accurately track the thermal history of dark atoms and the corresponding formation rate of dark molecules, including ionizing backgrounds from dark stars. However, such a calculation is beyond the scope of this paper and will be left for future work.

Molecular cooling proceeds through two mechanisms, rotational cooling and roto-vibrational cooling. Both mechanisms are quadrupole-suppressed for $H_{2}$. The Einstein $A$ coefficients (rate of transitions) for both the processes are given by \cite{Ryan:2021dis}
\begin{align}
    \label{eq:Amol}
    \Gamma_{\rm{rot}} &\simeq \alpha^{7}_{D} \frac{m^{6}_{c}}{m^{5}_{X}}\nonumber,\\
    \Gamma_{\rm{ro-vib}} &\simeq \alpha^{7}_{D} \frac{m^{7/2}_{c}}{m^{5/2}_{X}}\,.
\end{align}
We will assume the rates above to be equal to the quoted RHS in what follows \footnote{Ideally the $\mathcal{O}(1)$ factor of the rates above can be deduced by comparing to SM $H_{2}$ data. But for our order of magnitude estimates, this does not have a large effect.}.
Note that molecular cooling is quadrupole suppressed as well as suppressed by powers of $\frac{m_{c}}{m_{X}}$ and is not always efficient. 
The parameter space in which dark atoms can cool efficiently will eventually evolve to form dark stars. In the next section we provide details on the calculations needed to calculate the mass of dark stars.

\section{Mass of Dark Stars -- Jeans Mass}
\label{sec:mj}
In this section we will discuss the collapse of a dark gas clump into a black hole. Our method of calculating the mass of the dark star will be similar to that for SM baryons. If the diffuse gas  cools efficiently, it is continually described by an isothermal distribution. The gas progressively loses pressure support and collapses to higher densities. As the density increases, sound waves travel to increasingly smaller distances within a free-fall time $t_{\rm{dyn}}$ and encompass a smaller mass fraction of the dark clump. Thus the dark gas clump fragments to larger densities (smaller Jeans mass) until cooling becomes increasingly inefficient at higher densities. The Jeans radius and mass are given by
\begin{align}
    \label{eq:Jeans}
    R_{J} &= \left(\frac{15 T}{4 \pi G m^{2}_{X} n_{X}}\right)^{1/2} \nonumber, \\
    M_{J} &= \frac{4 \pi}{3} m_{X} n_{X} R^{3}_{J}.
\end{align}
Note that the gas fragments to smaller Jeans mass as long as the gas temperature $T$ scales slower than $n^{1/3}_{X}$. In the absence of nuclear processes, the final mass of the star, and hence the mass of the remnant black hole, is likely to be close to the final Jeans mass, absent significant mass loss mechanisms (i.e. we assume mass of dark star to be equal to the mass of the black hole $M_{*} = M_{\rm BH}$ ). The final Jeans mass is set by the temperature and density of the gas when cooling becomes inefficient. This can happen if the gas reaches local thermal equilibrium or if the cooling becomes opacity limited (i.e. the photons get absorbed well before they can escape the gas cloud). We will discuss the Jeans mass for the cases of atomic cooling and molecular cooling, detailing the area where we get $M_{\rm BH} = M_{J} = 100 \Msol$. We choose this value because it is squarely within the BHMG, and our constraints are calculated assuming a $100 M_{\odot}$ dark star. It should be noted that the following calculation of the Jeans mass serves as an order of magnitude estimate at an $\mathcal{O}(10)$ level due to the uncertainties associated with the Jeans mass calculation, the inexact correspondence between $M_{J}$ and $M_{\rm BH}$ and ignoring accretion that can increase the mass of the BH. Thus we consider $30-300 M_{\odot}$ Jeans mass as the range where we can reasonably expect to create a $100 M_{\odot}$ BH. 

\subsection{Atomic Jeans Mass}
As noted above, the Jeans mass is set by the temperature and density of the gas when cooling becomes inefficient. Cooling can become inefficient if the gas reaches thermal equilibrium. However for atomic cooling the timescale for the lowest-energy transition  $(2p \rightarrow 1s)$ is extremely short, making the density at which thermal equilibrium is reached very large. Cooling also becomes inefficient if the opacity destroys photons escaping the gas cloud, a phenomenon which transitions the cloud from efficient volume cooling to inefficient surface cooling. In all of the parameter space we consider, opacity is set by free-free absorption (inverse process of bremsstrahlung) \footnote{The atomic transition from ($2p \rightarrow 1s$) is fast and opacity due to this transition is only effective at very large densities and is not relevant to the Jeans mass calculation for our parameter space.}. The Rosseland mean opacity of an ionized gas for free-free absorption is given by 
\begin{align}
    \label{eq:Rosseland}
    \alpha_{R,ff} &= 0.38 \times \frac{\alpha^{3}_{D} n^{2}_{X}}{m^{3/2}_{c} T^{7/2}} \nonumber\\
    &= 1.3\times 10^{4} \times \frac{n^{2}_{X}}{m_{c}^{5}\alpha^{4}_{D}}\,,
\end{align}
where in the last line we have assumed that the gas can cool down efficiently to $T = \frac{B_{X}}{10}$, where $B_{X} = \frac{\alpha^{2}_{D} m_{c}}{2}$ is the binding energy of dark atoms. We approximate the Jeans mass by setting $\alpha_{R,ff} R_{J} = 1$, i.e. the mean free path of the photons is the size of the gas cloud. Using Eqs.~\ref{eq:Jeans} we get

\begin{align}
    \label{eq:JeansAtomic}
    M_{J} \simeq 200 \Msol \left(\frac{\alpha_{D}}{\alpha}\right)^{2} \left(\frac{m_{p}}{m_{X}}\right)^{7/3}\,.
\end{align}
Note that if the cooling is atomic then the Jeans mass is independent of the electron mass. If we had assumed that the gas can cool down efficiently till $B_{X}$ instead of $0.1 B_{X}$ we would have gotten Jeans mass higher by a factor of $\approx 3$ reflecting the uncertainties associated with calculating the Jean's mass. Thus as mentioned above the Jeans calculation is valid to $\mathcal{O}(10)$ level. In the plots below we show the region that accounts for black holes with mass $M \in (30 \Msol - 300 \Msol)$. This target range is shown in Fig.~\ref{fig:atomicstar} (blue band).

\subsection{Molecular Jeans Mass}
\label{sec:molMJ}

Molecular cooling becomes inefficient once the gas reaches local thermal equilibrium (LTE). Since the quadropole transition timescales for molecular cooling are much larger compared to atomic cooling, the gas reaches LTE at relatively low densities.
The density at which LTE is achieved can be calculated by equating the scattering rate $(n_{LTE}\sigma v)$ (where $v \simeq \sqrt{T/m_{X}}$) to the transition rate of $H_{2}$ molecules given in Eq.~\ref{eq:Amol}. For the purposes of our calculation we will assume a geometric cross section $\sigma = \pi a^{2}_{0}$ where $a_0$ is the Bohr radius. For a more detailed calculation see Ref.~\cite{Ryan:2021dis}.
The final temperature that the gas reaches depends on the dominant cooling mechanism and is given by
\begin{align}
    \label{eq:Tfinmol}
    T_{f,rot} &= \frac{\alpha^{2}_{D} m^{2}_{c}}{m_{X}}, \nonumber\\
    T_{f,ro-vib} &=  \frac{\alpha^{2}_{D} m^{3/2}_{c}}{m^{1/2}_{X}}\,.
\end{align}

Then using Eq.~\ref{eq:Amol} and Eq.~\ref{eq:Tfinmol}, and setting the collisional timescale equal to the molecular de-excitation timescale we get

\begin{align}
    \label{eq:LTEmol}
    n_{LTE,rot} = 10^{3} {~\rm cm^{-3}} \left(\frac{\alpha_{D}}{\alpha}\right)^{8}\left(\frac{m_{c}}{m_{e}}\right)^{7} \left(\frac{m_{p}}{m_{X}}\right)^{4},\nonumber\\
    n_{LTE,ro-vib} = 2\times 10^{8} {~\rm cm^{-3}} \left(\frac{\alpha_{D}}{\alpha}\right)^{8}\left(\frac{m_{c}}{m_{e}}\right)^{19/4} \left(\frac{m_{p}}{m_{X}}\right)^{7/4}\,,
\end{align}

where we have fixed the coefficient to match the SM values for $n_{LTE,rot}$ in \cite{bromm_formation_2002}. Then using Eq.~\ref{eq:Jeans} and Eq.~\ref{eq:LTEmol} we get that

\begin{align}
    \label{eq:MJmol}
    M_{J,rot} &\simeq 5\times 10^{3} \Msol \left(\frac{m_{e}}{m_{c}}\right)^{1/2} \left(\frac{\alpha}{\alpha_{D}}\right) \left(\frac{m_{p}}{m_{X}}\right)^{3/2} \nonumber,\\
    M_{J,ro-vib} &\simeq 3\times 10^{3} \Msol \left(\frac{m_{e}}{m_{c}}\right)^{1/8} \left(\frac{\alpha}{\alpha_{D}}\right) \left(\frac{m_{p}}{m_{X}}\right)^{15/8}\,.
\end{align}
The gas cloud will fragment within dynamical time to the smallest allowed Jeans mass. For $m_{X} = 1\GeV$, comparing the Jeans molecular mass Eq.~\ref{eq:MJmol} to the atomic Jeans mass Eq.~\ref{eq:JeansAtomic}, we can show that the atomic Jeans mass is the lowest in our desired target range of $30 - 300 \Msol$ and our allowed parameter space. For $m_{X} = 10\GeV$, the minimum between atomic and molecular Jean's mass is lower than $30 \Msol$. Thus the only way we get black holes in the target range of $30 - 300 \Msol$ is assuming molecular cooling is inefficient.  Thus we only consider the atomic Jeans mass in Eq~\ref{eq:JeansAtomic}.

\section{Stellar Evolution}
\label{sec:evolution}

The dark stars form with masses given by the Jeans mass calculated in the above section. Under the assumption that the dark gas clump is in hydrostatic equilibrium, we can use the stellar equations to describe the evolution of the system. Since we assume no nuclear physics in the dark sector, the dark stars are purely gravity powered, i.e. the gravitational potential energy lost due to contraction is radiated away by photons. We assume nuclear reactions are absent in the dark sector. To begin with we also assume that there is no neutrino-like energy loss mechanism. However, as we will see, some additional energy loss mechanism (see e.g. \cite{Croon:2020ehi,Croon:2020oga}) or the presence of heavier dark elements is needed if we want a star with $m_{X} = 1\GeV$ to collapse into a BH within the mass gap before $z=2$. 

\begin{figure}
    \centering
    \includegraphics[width=0.495\columnwidth]{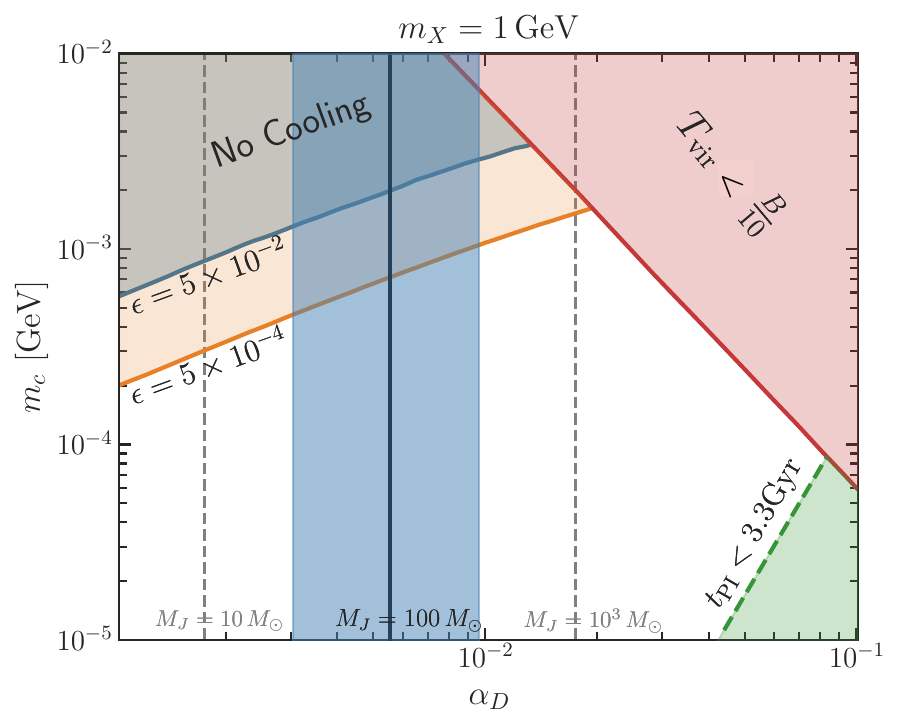}
    \includegraphics[width=0.495\columnwidth]{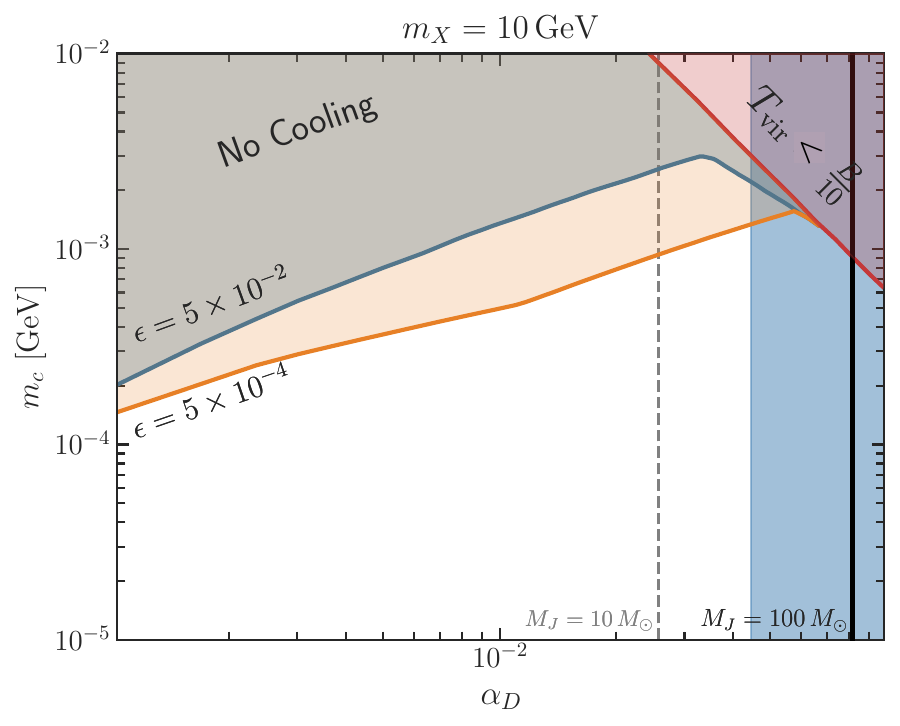}
    \caption{\textbf{Left:} Target region where we can expect $100 M_{\odot}$ BH (blue band) and constraint for core temperature of star to reach pair instability temperatures before z = 2 (3.3 {\rm Gyr}) (green region) for $m_{X}=1\ \GeV$. \textbf{Right:}  Target region where we can expect $100 M_{\odot}$ BH (blue band) for $m_{X} = 10 \GeV$. The $m_{X} = 10\ \GeV$ case is not affected by this constraint coming from time to reach pair instability.}
    \label{fig:atomicstar}
\end{figure}

The absence of nuclear reactions makes it possible to study the evolution of dark stars analytically. In order to understand the stellar evolution and possible collapse into a BH, we need to describe the central density $\rho_{c}$ as well as central temperature $T_{c}$ of the star. Once  $T_{c}$ becomes comparable to the mass of the dark electron $m_{c}$, pair instability can set in, triggering the collapse of the star. Since stellar cooling is due to radiation from the surface, we also need to know the surface temperature $T_{s}$, so we can confirm that the star can collapse till the central temperature reaches the pair instability region by $z = 2$. We can examine $\rho_{c},\ T_{c}$ and $T_{s}$ by modelling the star as a polytrope. A polytropic profile implies a global relation between pressure and density of the form $P = \rho^\gamma$, where $\gamma = (n+1)/n$. This simple treatment for solving systems in hydrostatic equilibrium is a surprisingly powerful tool for understanding stellar properties. In particular, massive stars are very close to a $n = 3$ polytrope. This is because they are primarily supported by radiation pressure and ion pressure. Assuming that relative contribution from each, parameterized here as $\beta \equiv P_{\rm{gas}}/P$, is constant throughout the star, the total pressure scales as $P \sim \rho^{4/3}$, corresponding to a polytrope with $ n = 3$. We find reasonably good agreement between the $T_c$, $T_s$, and $\rho_c$ values calculated using the $n = 3$ polytrope and those calculated using MESA (see Figure \ref{fig:Tsurf100} and Appendix \ref{sec:appcollapse} for details). 
The central density and temperature of the star evolve as
$$T_{c} = \xi \rho^{1/3}_{c} \,,$$
where $\xi= \left(\frac{45 (1-\beta)}{\pi^{2} (\mu m_{X}) \beta}\right)^{1/3}$, and $\mu$ is the mean molecular weight.
The derivation of $T_{c},\xi, T_{s}$ as well as the derivation of the mass of the stars that reach the pair instability region is given in App. \ref{sec:appcollapse}. Here we just quote the results.
The  potential energy of the dark star that needs to be radiated away for a $n = 3$ polytrope is $E_{*} = \frac{3}{2} \frac{G M^{2}_{*}}{R}$ where $R$ is the radius of the star. The temperature of the star is calculated in Eq.~\ref{eq:Ts} and is given by

\begin{multline}
    \label{eq:Tsnat}
    T_{s} = 9.7\times 10^{-9} \GeV \left(\frac{100}{f}\right)^{2/7}\left(\frac{\alpha}{\alpha_{D}}\right)^{6/7} \left(\frac{m_{c}}{m_{e}}\right)^{3/7}\left(\frac{m_{X}}{m_{p}}\right)^{2/21} \\
    \times \left(\frac{(1-\beta)^{13}}{\beta^{10}}\right)^{1/21} \left(\frac{\rho_{c}}{10^{-15} \GeV^{4}}\right)^{4/21} \,,
\end{multline}
where $f$ is the ratio of bound-free to free-free opacity (see App.\,\ref{sec:appcollapse}).
Thus the cooling time, defined as the time it takes the star to reach temperatures where pair instability can set in i.e. $T_{c} = m_{c}/10$, is given by 

\begin{align}
t_{\rm{PI}} = \frac{E*}{4 \pi T^{4}_{s} R^{2}} =  \frac{\frac{3 G M^{2}_{*}}{2R}}{4 \pi T^{4}_{s} R^{2}}\,.
\end{align}
The cooling time scales as $t_{PI} \propto \rho_{c}^{5/21}$ i.e. it gets harder to cool the star as central density increases. Thus, majority of the total evolution time spent reaching $T_{c}$ is dominated by the period with the highest temperature, $T \simeq T_{c} \simeq m_{c}/10$ (i.e. $\rho_{c} = \left(T_{c}/\xi\right)^{3}$), after which pair instability can occur. The time until pair instability sets in is given by

\begin{align}
\label{eq:tPI}
t_{\rm{PI}} = 5.3\times 10^{4} \, \mathrm{yrs}  \left(\frac{f}{100}\right)^{8/7} \left(\frac{\alpha_{D}}{\alpha}\right)^{24/7} \left(\frac{m_{e}}{m_{c}}\right) \left(\frac{m_{X}}{m_{p}}\right)^{13/7} \left(\frac{M_{*}}{100 \Msol}\right)^{2} \left(\frac{\beta^{29/7}}{(1-\beta)^{45/14}}\right)\,.
\end{align}
Note that $\beta$ also has a dependence on $m_{X}$ and $M_{*}$ (Eq.~\ref{eq:beta}). In principle we need to evaluate $t_{\rm{PI}}$ assuming the star mass to be the Jeans mass i.e. $M_{*} = M_{J}$. However to denote our constraint, we will use $M_{*} = 100 \Msol$ i.e. a value inside the BHMG. In Fig.~\ref{fig:atomicstar} we show the associated constraint from $t_{\rm{PI}} < 3.3 \mathrm{Gyr} ~ (z=2)$ (green region). The constraint should be treated as approximate since it's a near linear function of our approximate factor $f$. This constraint does not rule out our region of interest of atomic Jean's mass between $30-300 M_{\odot}$ for $m_{X} = 1\GeV$. and does not affect any parameter space for $m_{X} = 10 \GeV$. Additional cooling mechanisms in the dark atomic sector can significantly reduce $t_{\rm{PI}}$, however we don't consider them here.
Even when $T_{c}$ reaches temperatures where pair production occurs, collapse through pair instability is not guaranteed. The question of dynamical stability through pair-production motivates introducing the first adiabatic index, given by

\begin{equation}
\label{eq:gamma1}
    \Gamma_1 = \frac{\rho}{P} \Big( \frac{\partial P}{\partial \rho}\Big)_s\,,
\end{equation}
the logarithmic derivative of pressure with respect to density at fixed entropy. It can be shown, assuming homologous and adiabatic compression, that dynamical stability requires $\Gamma_1 \gtrsim 4/3$. This can be seen as follows: the pressure on a unit area of a concentric sphere in hydrostatic equilibrium is given by

\begin{equation}
    P = \int_m^M \frac{Gm}{4\pi r^4}dm \, .
\end{equation}
If we compress the star adiabatically and assume homology (self-similarity of stellar evolution at all radii), then a radial shell being compressed from $R$ to $R'$ will change the right hand side as $\sim \Big(\frac{R'}{R}\Big)^{-4}$, whereas the pressure will scale as $\Big(\frac{\rho'}{\rho}\Big)^{\Gamma_1} \sim \Big(\frac{R'}{R}\Big)^{-3 \Gamma_1}$. Thus, if $\Gamma_1 \lesssim 4/3$, the weight on the sphere will increase faster than the counteracting pressure and the sphere will collapse. Note that this condition is approximate because small corrections may be present due to relativistic effects. However, these  corrections are typically only relevant for neutron stars, hence why we will assume a critical value of $\Gamma_1 = 4/3$. 
%%%
Here we consider $M_{\rm{crit}}$ to be the lowest mass for which the core experiences pair instability. We have solved for this instability in App. \ref{sec:appcollapse} numerically with the result \footnote{Ideally, the collapse condition should be that the average $\Gamma_{1}$ over the entire star is below $4/3$. However, we only consider the local value of $\Gamma_{1}$ in the core. Thus $M_{\rm{crit}}$ derived here is likely smaller than the true value.}

\begin{align}
    \label{eq:Mcrit}
    M_{\rm{crit}} \simeq 260 \Msol \frac{1 \GeV}{m_{X}}\,.
\end{align}
Note that this gives a lower bound on the mass of the dark proton, if we require the dark stars to form black holes in the mass gap. For $M_{J} \gtrsim M_{\rm{crit}}$ the star will experience pair instability and collapse. Note that most of the region for $m_{X} = 1\GeV$ does not experience pair instability for a $M_{*} = 100 \Msol$ and hence will not collapse into a BH within $3.3~ \rm{Gyr}$. Collapse can be achieved by assuming additional cooling mechanisms or the presence of a significant fraction of  ``dark helium'' or heavier ``dark elements'', at the expense of minimality of the dark sector.

\section{Constraints and Results}
\label{sec:constraints}

If dark atoms make up a fraction of dark matter, it is subject to several constraints from early universe (CMB and matter power spectrum) to late universe astrophysics (galaxy morphology, ultra-faint dwarfs, lensing etc.). Most of these constraints are independent of whether  the dark atoms collapse eventually into black holes.

\subsection{CMB and $N_{eff}$}
Dark atoms undergo dark acoustic oscillations, analogous to the baryons in SM and the dark photons contribute to $N_{eff}$. These constraints have been derived in \cite{Cyr-Racine:2012tfp,Cyr-Racine:2013fsa} and the parameter space we consider here already satisfies these constraints by assuming $\epsilon \leq 0.05$ and $r = \frac{T_{d,0}}{T_{0}} \leq 0.5$.

\subsection{Galaxy Morphology}
\begin{figure}
    \centering
    \includegraphics[width=0.495\columnwidth]{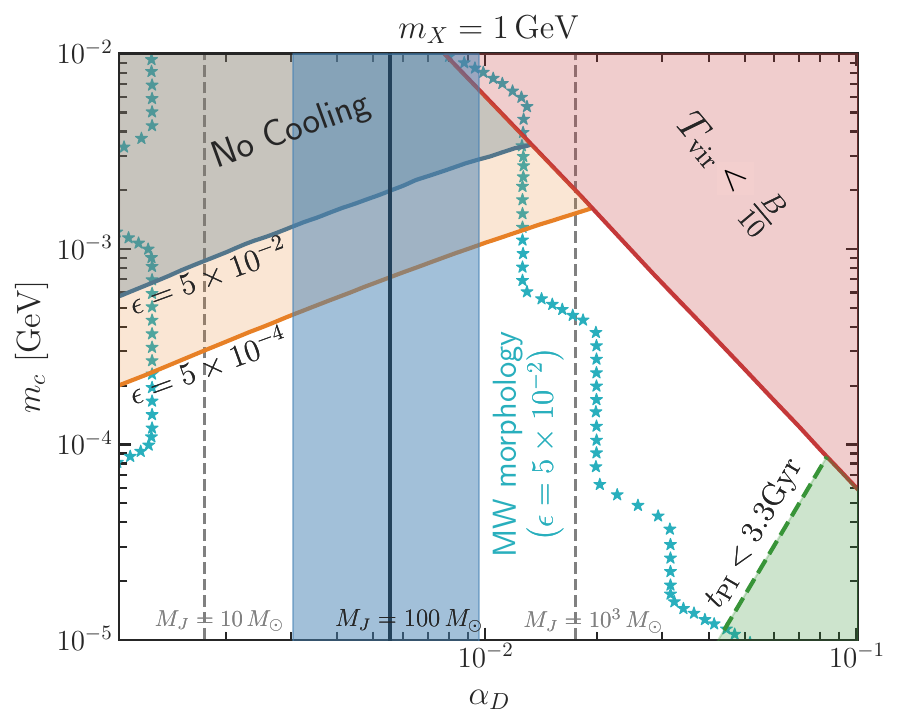}
    \includegraphics[width=0.495\columnwidth]{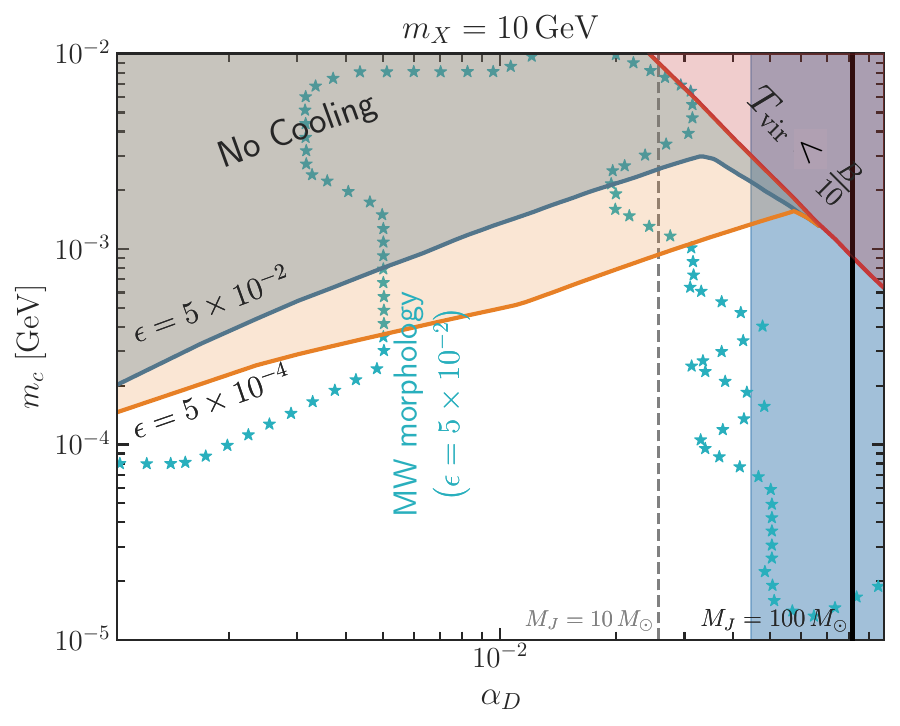}
    \caption{\textbf{Left:} Constraint plots including the Galaxy mass and morphology constraints (starred blue) from \cite{Ghalsasi:2017jna} for $m_{X}=1\GeV$ and $\epsilon = 0.05$. \textbf{Right:} Constraint plots including the Galaxy mass and morphology constraints (starred blue) from \cite{Ghalsasi:2017jna} for $m_{X}=10 \GeV$ and $\epsilon = 0.05$. Note that the ``discretized'' nature of the constrains comes from limited merger tree simulation data in the $\alpha_{D},m_{c}$ plane.}
    \label{fig:final}
\end{figure}

The primary constraint on the fraction of dark matter constituted by dark atoms comes from the morphology of dark galaxies. For $\epsilon = 0.05$, dark atoms typically form a bulge-like galaxy at the center of the Milky Way \cite{Ghalsasi:2017jna}. The mass of our galaxy is well measured with stellar rotation curves as well as from measurements of stellar luminosity, and the two measurements agree up to a factor of 2. Thus, atomic dark matter in the bulge cannot contribute more than the mass of the stars in the center of the Milky Way. This gives the constraint \cite{Ghalsasi:2017jna}
\begin{align}
    \label{eq:morph}
    \epsilon \times  f_{bulge} \left(\frac{r_{MW}}{r_{s}}\right)^{3} < 0.02 \,,
\end{align}
where $r_{MW} = 3\ \rm{kpc}$ is the stellar radius of the Milky Way and $r_{s}$ is the radius of the dark galaxy. For dark galaxies which form a bulge i.e. $f_{bulge} \simeq 1$ instead of the disk $r_{s} \ll r_{MW}$ providing a stringent constraint of $\epsilon$. Note that $r_{s}$ is a function of $\epsilon, \alpha_{D}, m_{c}, m_{X}$ since it dictates the merger history and hence morphology of the dark galaxy. For $\epsilon = 0.05$, a large part of the parameter space is ruled out and the constraints derived in \cite{Ghalsasi:2017jna} are plotted in Fig.~\ref{fig:final} (dotted brown) \footnote{The discretized nature of the constraint is a limitation arising from the fact that merger tree simulations have been performed for discretized grid in $\alpha_{D},m_{c}$ parameter space.}. It is possible that $\epsilon =  5\times 10^{-4}$ evades this constraint, however an independent set of merger tree simulations need to be performed with $\epsilon = 5\times 10^{-4}$. We leave this for future work.
If the dark atoms form a disk instead of a bulge, an equivalent constraint of $\epsilon < 0.05$ comes from {\it Gaia} measurements of the Milky Way surface density \cite{Schutz:2017tfp}. However since most of the parameter space for $\epsilon = 0.05$ does not form a disk, we do not plot these constraints here.

\subsection{Other Constraints}
We now discuss constraints that don't affect our scenario or require additional modelling in order to derive them.

\textbf{Dynamical Friction} causes the heavier dark stars to sink to the center of dwarf galaxies, imparting their momentum to SM stars resulting in ``dynamical heating" of SM stars and increasing the half light radii of the ultra faint dwarfs \cite{Brandt:2016aco}. This gives the constraint $$\epsilon \times f_{cooled} \times M_{*} < 10 \Msol \,,$$ where $f_{cooled}$ is the fraction of gas in the halo that has cooled and formed stars. However since we consider $\epsilon < 0.05$ and $M_{*} \simeq 100 \Msol$ in the mass gap, this constrain is not applicable to us.

\textbf{Microlensing} surveys like MACHO \cite{MACHO:2000qbb} and OGLE \cite{2015ApJS..216...12W} typically put strong constraints on sub solar mass black holes and are not relevant for our target parameter space of $M_{*}\simeq 100 \Msol$. See \cite{Ghalsasi:2017jna} for further details.

Presence of dark atoms inside a SM star modify the \textbf{stellar mass luminosity relation} \cite{Peled:2022byr}. These constrain the mass fraction of dark atoms inside star to be $\lesssim 5\%$. However since it's not straightforward to translate the total fraction of dark matter that is atomic $(\epsilon)$ to the fraction that ends up in SM stars we won't consider this constraint.

Similarly, the presence of \textbf{baryons inside a dark star} can also lead to observable signatures \cite{Howe:2021neq}. However, these constraints depend on the kinetic mixing between dark and SM photon, which we have ignored here.

\section{Discussion, Conclusions, and Future Work}
\label{sec:conclusions}

Observations made by LIGO, Virgo and KAGRA collaboration of merging binary black holes with masses within the BHMG are in direct conflict with our understanding of standard model stars. Electron-positron pair instability within the cores of SM stars reduces the supportive radiation pressure and leads to a runaway collapse and subsequent thermonuclear burning of oxygen, causing a destructive explosion that causes a star to shed mass or leaves no remnant black hole at all.
This motivates a scenario in which dark analogues of the electron, proton, and photon are able to dissipate energy through radiation but are unable to generate energy through fusion. In the absence of nuclear reactions, the radiated energy of a \textit{dark} star is sourced by gravitational collapse alone and runaway collapse can directly form a black hole from the compressed dark stellar matter. In this case, the instability of a dark star does not lead to a violent explosion due to a pair instability supernova (PISN). Thus, there is no analogous BHMG for dark stars. 
We analytically and numerically explored the best-motivated regions of parameter space to determine the conditions in which black holes would form in the mass gap. We showed that under reasonable assumptions about the dominant atomic and molecular cooling mechanisms, the Jeans mass of dark stars can fall within the BHMG. Furthermore, assuming a polytropic model and comparing with state-of-the-art stellar evolution code, we demonstrated that dark stars can reach a stage of dark electron-positron production so long as $m_{X} \simeq 10\ \GeV$, leading to dynamic instability and collapse.  With no fusion to support these dark stars, they can naturally form black holes with masses in the BHMG before $z = 2$, allowing time for these black holes to merge and be detectable by LIGO, Virgo and KAGRA collaboration. Intriguingly, we discovered that the black holes that form from dark sector collapse would not have mass within the mass gap if the dark sector proton were lighter than the visible sector proton.

While we present a potential explanation for black holes in the BHMG, further work remains to be done to associate the black holes in the mass gap found by LIGO, Virgo and KAGRA collaboration to black holes formed by dark atoms. Since the density and merger rates of such black holes, as well as constraints on dark atoms, depend strongly on the dark galaxy morphology, it is essential to perform merger tree simulations similar to \cite{Ghalsasi:2017jna} for a larger grid of $m_{X},m_{c},\alpha_{D}, \epsilon$, and $\xi$. Specifically, improvements need to be made to the merger tree simulations by including molecular cooling, which could be important in the large $m_{c} ,\alpha_{D}$ parameter space. Effects of stellar feedback could also be important in the small $m_{c} ,\alpha_{D}$ parameter space. Furthermore, it is essential to compare $N$-body simulations to merger tree simulations in order to have a more accurate understanding of the galactic astrophysics of atomic dark matter. For simplicity, we assumed in this work that the mass of the remnant black hole formed from dark star collapse is approximately equal to the Jeans mass of the star. This is because there is no supernova expelling mass from the system. However, a full $N$-body simulation and/or thorough accretion modeling would be needed in order to more accurately determine the black hole's final mass. Modifying existing stellar evolution code, such as MESA, to include different $\alpha_{D}, m_{c}$ and $m_{X}$ will also let us compare our analytical estimates of stellar evolution to the full numerical solution. If there exists an appreciable population of black holes formed from the collapse of dark stars, and if one can predict the morphology of the dark galaxies in which the black holes reside, one can predict the merger rate of black holes in the mass gap. Current and future observations from LIGO, Virgo and KAGRA collaboration can then be used to place limits on the abundance of black holes formed through this mechanism and consequently on the dark atom parameter space.
    
Finally one may imagine adding nuclear physics to the dark sector. In this case, there may exist a dark BHMG with the upper and lower limits being determined by physics of dark nuclear reactions. It is possible that for different values of $m_X$ and $m_c$, there is a dark BHMG that does not fully or even partially overlap with that of the standard model. In this case, black holes could form over a full continuum of masses with some being created by SM stars and some created by dark stars. While we have focused in this work on the mass range relevant for the pair-instability mass gap of the standard model, dark stars could form black holes with much larger masses. In particular, this mechanism could contribute to seeding the supermassive black holes that are found at the center of large galaxies. On the other end of the spectrum, since the Chandrasekhar mass limit scales as $\sim 1/m_X^2$, it is possible that dark stars formed of atomic dark matter could also form sub-solar mass black holes, as envisioned in Ref.~\cite{Shandera:2018xkn}.

%%%%%%%%%%%%%%%%%%%%%%%%%%%%%%%%%%%%%%%%%%%%%%%%%%%%%
%%%%%%%%%%%%%%%%%%%%%%%%%%%%%%%%%%%%%%%%%%%%%%%%%%%%%
\acknowledgments
%%%%%%%%%%%%%%%%%%%%%%%%%%%%%%%%%%%%%%%%%%%%%%%%%%%%%
%%%%%%%%%%%%%%%%%%%%%%%%%%%%%%%%%%%%%%%%%%%%%%%%%%%%%
We are very grateful to Hiren Patel for his help in the early stages of this project. We also thank Matt McQuinn for many useful conversations. NS and AG would like to thank Stan Woosley for  help with questions in stellar evolution. NS and LS would  like to thank the MESA community for helping with MESA modifications as well as troubleshooting. AG, NF and NS would also like to thank the attendees of ``Workshop on Atomic Dar Matter'' for helpful conversations.  This material is based upon work supported in part
by the National Science Foundation Graduate Research
Fellowship under Grant No. DGE-1842400 to NS. This work is partly supported by the U.S.\ Department of Energy grant number de-sc0010107 (SP).The work of AG is supported by the U.S. Department of Energy under grant No. DE–SC0007914.  The work of NF is supported in part by DOE CAREER grant DESC0017840. NF would like to thank the Aspen Center for Physics, which is supported by National Science Foundation grant PHY-1607611, where this work was partly performed. LS received funding through the Ron Ruby Scholarship at UC Santa Cruz. 

%%%%%%%%%%%%%%%%%%%%%%%
\appendix
\section{Cooling times}
\label{sec:appcoolingtimes}
%%%%%%%%%%%%%%%%%%%%%%%

\subsection{Dark Atomic Cooling}
After dark recombination, the atomic dark matter falls into the gravitational potential wells created by the rest of the non-atomic DM halo during formation of early galaxies and virializes. If the virial temperature is large enough to ionize the gas or excite atomic transitions, then the atomic dark matter will cool through bremsstrahlung, compton scattering, and atomic transitions. The timescales for each of these processes are given by
\begin{align}
\label{eq:tcooling}
    t_{\rm brem} &= \frac{9}{2^{5}} \left(\frac{3\pi}{2}\right)^{1/2} \left(\frac{m^{3/2}_{c}T^{1/2}}{\alpha^{3}_{D}n_{c}(z)x^{2}}\right) \,,\\
    t_{\rm comp} &= \frac{135}{64\pi^{2} x} \left(\frac{m^{3}_{c}}{\alpha^{2}_{D}\left(T_{d0}(1+z)\right)^{4}}\right)\,,\\
    t_{\rm atomic}(T \gtrsim B_{X}/10) & = \frac{9}{2^{5}} \left(\frac{3\pi}{2}\right)^{1/2} \left(\frac{T^{2}}{\alpha^{6}_{D}n(z)}\right),
\end{align}
where $x$ is the ionization fraction, $T$ is the gas temperature, $n$ is the atomic DM number density and $T_{d0}$ is the dark CMB temperature today. The atomic cooling timescale here is only referenced for $T \gtrsim B_{X}/10$, a more general formula is given in Appendix A of \cite{Ghalsasi:2017jna}. The number density at  redshift $z$ is given by
\begin{align}
    \label{eq:nc}
    n(z) = \frac{\epsilon \rho_{m,0} \, \Delta (1+z)^{3} }{m_{X}},
\end{align}
where $\rho_{m,0}$ is the dark matter density today, $\epsilon$ is the fraction of DM made up of atomic dark matter, and $\Delta \approx 180$ is the overdensity associated with a virialized halo compared to the background dark matter density.
The transfer of energy (and hence the thermal equilibrium between the dark protons and electrons) is controlled by the timescale associated with particle-particle collisions and is given by

\begin{align}
    \label{eq:tequil}
    t_{\rm eq} = \frac{m_{X} m_{c}}{2\sqrt{3}\alpha^{2}_{D}n}\left(\frac{3T}{m_{c}}\right)^{3/2} \left(\log\left(1+\frac{T^{2}}{4\alpha^{2}_{D}n^{2/3}}\right)\right)^{-1}\,.
\end{align}
The maximum temperature that the gas reaches is the virial temperature of the halo given by
\begin{align}
    \label{eq:Tvir}
    T_{\vir} = \frac{1}{3} \left(\frac{4\pi}{3}\right)^{1/3} G M^{2/3}_{halo} m^{4/3}_{X} \epsilon^{-1/3} n(z)^{1/3}\,.
\end{align}
Note that the virial temperature is independent of $\epsilon$ but scales as $T_{\vir} \propto m_{X}$. Thus the equilibration timescale goes as $t_{\rm eq} \propto m^{7/2}_{X}$ for the halo of the same mass. Atomic dark matter $m_{X} \gg 10 \hspace{1mm} \GeV$ cannot cool in equilibrium, and we therefore ignore that possibility. Note that for $T > B_{X} $, $\frac{t_{\rm{atomic}}}{t_{\rm{brem}}} \simeq \left(\frac{T}{B_{X}}\right)^{3/2}$ and thus atomic cooling is more efficient for $T \approx B_X/10$, with bremsstrahlung becoming more efficient for $T \gg B_{X}$.
We will require that a significant fraction of the gas is able to virialize and cool by $z = 2$\footnote{The BH mergers for BH in the mass gap are estimated to be at $z\approx 1$. However since the gas needs to cool, form stars, evolve to collapse into a BH and then undergo a binary merger. Thus we take the benchmark of $z = 2$ for our calculations.}.  We require that $T_{\vir} > \frac{B_{X}}{10}$ for the majority of dark atoms. Imposing the above condition and using Eq.~\ref{eq:Tvir} we get

\begin{align}
    \label{eq:mhaloconst}
    M_{\rm halo} &\geq 2\times 10^{20} \left(\frac{m_{c}}{m_{X}}\right)^{3/2} \alpha^{3}_{D} \Msol  = 5.7\times 10^{20} \left(\frac{B_{X}}{m_{X}}\right)^{3/2} M_{\odot}\,.
\end{align}
The cumulative mass fraction of halos above a given mass is given by \cite{1974ApJ...187..425P,1991ApJ...379..440B,Lacey:1993iv}
\begin{align}
    \label{eq:cmf}
    P(>M,z=2) = \mathrm{erfc}\left(\frac{\delta_{c}(z = 2)}{\sqrt{2}\sigma(M)}\right) \geq 0.1\,,
\end{align}
where we assume that at least $10\%$ of the dark matter should be in halos that have temperatures higher than $B_{X}/10$ and hence can potentially cool. Note that the constraints we will derive are sensitive to the fractions we assume. Throughout, we use a conservative value of $10\%$.
Above, $\delta_{c}(z) = 1.686 \frac{gf(0)}{gf(z)}$, where $gf(z)$ is the growth function of matter perturbations given as
\begin{align}
    \label{eq:gf}
    g(z) = \frac{5}{2} \Omega_{m} \frac{H(z)}{100 h} \int_{0}^{1/(1+z)} \left(\frac{100 h}{a H(a)}\right)^{3} da \,,
\end{align}
and where $\sigma(M)$ is the standard deviation of  matter overdensities, given by

\begin{align}
    \label{eq:sigma}
    \sigma^{2}(M) = \int dk \frac{k^{2}}{2\pi^{2}} P(k)\left(\frac{j_{1}(k R)}{k R}\right)^{2},  
\end{align}
where $P(k)$ is the linear power spectrum at $z = 0$ and $j_{1}$ is the first spherical Bessel function and $R = \left(\frac{3M}{4\pi \rho_{m}}\right)^{1/3}$ is the Lagrange radius of a halo with mass $M$.
Using Eqs.~\ref{eq:cmf},~\ref{eq:gf},~\ref{eq:sigma} we get that
$$\sigma(M) = \frac{\delta_{c}(z = 2)}{\sqrt{2} \times 1.15} = 2.37, $$ where we have used $\Omega_{m} = 0.26, h = 0.7$ in the above calculation. Using Eq.~\ref{eq:sigma}, this corresponds to a halo mass of $M \simeq 3\times 10^{11} \Msol$. If the lower bound on halo mass given in Eq.~\ref{eq:mhaloconst} is larger than halo mass of $3 \times 10^{11} \Msol$ derived above, the fraction of gas that can cool efficiently is less than $10 \%$.
This gives us the constraint
\begin{align}
    \label{eq:coolingconst1}
    B_{X} =\frac{\alpha^{2}_{D} m_{c}}{2} < 5.4 \times 10^{-7} m_{X}\,,
\end{align}
which is plotted in Fig.~\ref{fig:tcool}.
The parameter space not bounded by the above constraint can cool efficiently as long as
$$  {\rm min} (t_{\rm{brehm}},t_{\rm{compton}},t_{\rm{atomic}}) < t_{\rm{dyn}} \qquad  \text{and} \qquad t_{equil} < t_{\rm{dyn}} $$ for a given halo. Here, $t_{\rm dyn} = \frac{1}{\sqrt{8\pi G \rho_{m}}}$ is the approximate timescale associated with doubling of the halo mass through mergers and accretion. Different processes are efficient at cooling the halos at different times and the exact fraction of halos that cool has to be understood through a merger tree \cite{Ghalsasi:2017jna} or a hydrodynamical simulation \cite{Roy:2023zar, Gemmell:2023trd}. Merger tree or hydrodynamical simulations will be necessary to model the distribution of black holes which is an important first step in understanding the merger rates of the black holes. We however do not attempt to predict the merger rates in this work. Thus for our purposes a semi-analytical understanding (described below) of the parameter space that can cool efficiently is sufficient.

From Eqs.~\ref{eq:tcooling} we understand that cooling is most efficient for high $z$ as long as the gas can be ionized i.e. $T \gtrsim B_{X}/10$. Cooling is also efficient for low virial temperature (except for Compton cooling which does not depend on the gas temperature). Thus once $T_{\vir} \approx B_{X}/10$, the gas can cool. Since virial temperature increases with decreasing redshift $z$, one can find a $z_{\rm max}$ for which at least $10\%$ of the gas has a temperature above $B_{X}/10$. Then we can compare cooling and equilibration times to the dynamical time of the halo to find out if cooling is efficient. Here is the algorithm we follow:
\begin{itemize}
    \item Solve $M(z)$ s.t. $P(>M,z) = 0.1$. This corresponds to solving
    \begin{align}
        \sigma(M(z)) = \frac{\delta_{c}(z)}{\sqrt{2} \times 1.15} \,,
    \end{align}
    
    \item with $M_{\rm halo} = M(z)$ solve for $z_{ \rm max}$ using
    \begin{align}
        T_{\vir}(z) = \frac{B_{X}}{10}\,.
    \end{align}

    \item Check if the cooling and equilibration times are smaller than the dynamical timescale of the halo.
\end{itemize}
Following the above steps, the parameter space that can cool efficiently is shown in Fig.~\ref{fig:tcool}.

\section{$n = 3$ polytrope evolution and collapse}
\label{sec:appcollapse}

In this section we will discuss the evolution of the star assuming the star is a $n=3$ polytrope. The $n=3$ polytropic solution will be applicable when the core temperature $T_{c} < m_{c}$ i.e. we will investigate the early stages of our star's evolution. In order for a polytropic solution to apply, we will assume that our dark star is in hydrostatic equilibrium. We will also assume that we are in a regime (the star is heavy enough) that the pressure inside the star is a combination of non-relativistic gas pressure and radiation (i.e. we avoid electron degeneracy pressure experienced by lighter stars closer to the Chandrasekhar mass)\footnote{The model described below is also known as the Eddington $\beta$ model in literature.}. 
Since the lower limit of the mass gap is much higher than the Chandrasekhar mass for $m_X \gtrsim 0.1 \GeV$, this assumption will hold for all cases studied here.
The pressure inside the star can be written as (note everything is in natural units with $\hbar = c = k_{B} =1$)
\begin{align}
    \label{eq:P}
    P &= P_{gas} + P_{rad}\\
       &= \frac{\rho}{\mu m_{X}}  T + \frac{\pi^{2}}{45} T^{4}\,,
\end{align}
where $\mu = 0.5$ is the mean molecular weight in units of $m_{X}$. 
Assuming $\beta = P_{gas}/P$ and $T = \xi \rho^{1/3}$ we get
\begin{align}
\label{eq:xi}
\xi = \left(\frac{45 (1-\beta)}{\pi^{2} (\mu m_{X}) \beta}\right)^{1/3} \,.
\end{align}
 The assumption of $T = \xi \rho^{1/3}$ then gives us a $n = 3$ polytrope with $P = K \rho^{4/3}$ with $$K  = \frac{\xi}{\beta (\mu m_{X})} = \left(\frac{45}{\pi^{2}}\right)^{1/3} \left(\frac{(1-\beta)}{(\beta \mu m_{X})^{4}}\right)^{1/3}\,.$$
 For an $n = 3$ polytrope we can define change of variable into a dimensionless parameter $z = A r $ where $$A = \left(\frac{\pi G}{K}\right)^{1/2} \rho_{c}^{1/3},$$ where $\rho_{c}$ is the central density of the star. The polytrope solution to the density is given by the Lane-Emden equation of n = 3 given by
\begin{align}
    \label{eq:LE}
    \frac{d^{2}w}{dz^{2}} + \frac{2}{z}\frac{dw}{dz} + w^{3} = 0\,.
\end{align}
The solution has the first zero at $z_{3} = 6.89$ which corresponds to the radius of the star $R = \frac{z_{3}}{A}$. The density at a given radius, parameterized by $w$, is given by $\rho = \rho_{c} w^{3}$. The mass of the star $M_{*}$ can be written as 
\begin{align}
    \label{eq:Mstar}
    M_{*} = \int_{0}^{R} 4\pi r^{2} \rho(r) dr = \int_{0}^{z_{3}} 4\pi \frac{\rho_{c}}{A^{3}} w^{3}(z) z^{2} dz   \simeq 1225.39 \times \frac{M^{3}_{\rm pl}}{(\mu m_{X})^{2}} \left(\frac{1-\beta}{\beta^{4}}\right)^{1/2}\,,
\end{align}
where $M_{\rm pl} = 2.435\times 10^{18}\ \GeV$ is the reduced Planck mass. Note that $\rho_c$ dependence disappears for $M_{*}$ for n = 3 polytrope. This uniquely determines $\beta$ in terms of $M_{*}$.
\begin{align}
    \label{eq:beta}
    \frac{\beta^{2}}{(1-\beta)^{1/2}} = 15\times \left(\frac{\Msol}{M_{*}}\right) \left(\frac{\GeV}{\mu m_{X}}\right)^{2}\,.
\end{align}

\subsection{Cooling timescale for $n = 3$ Polytrope}
\begin{figure}[t!]
    \centering
    \includegraphics[width=0.70\textwidth]{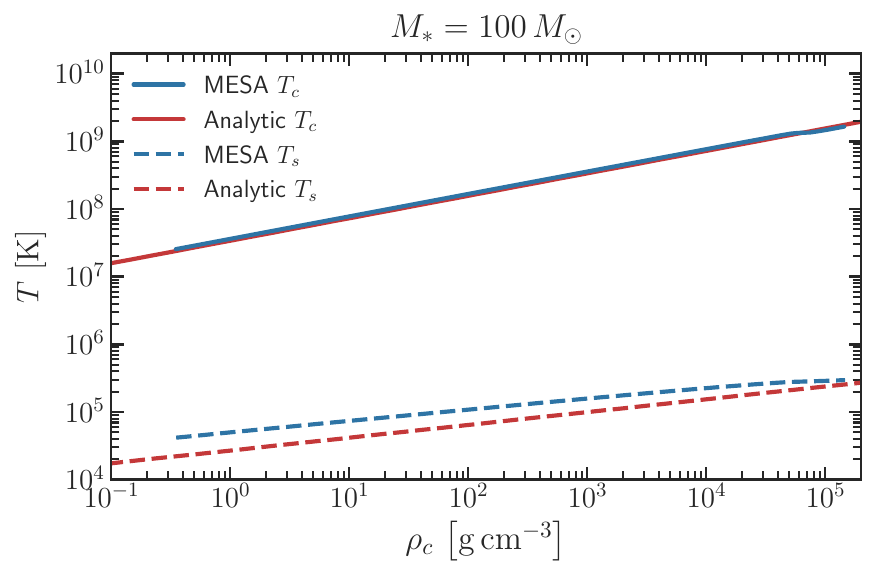}
    \caption{Comparison of temperature-density evolution between MESA (blue) and equations \ref{eq:xi} and \ref{eq:Ts} (red) for $M_{*}=100\Msol$. The dashed lines indicate a comparison between the surface temperatures, and the full lines represent the core temperatures.We have only shown the mesa evolution up to $T_{c} = m_{c}/10$.}
    \label{fig:Tsurf100}
\end{figure}

In order to figure out if the star cools within the lifetime of the universe we need to find out the surface temperature of the star. It's true that $T(R) = 0$ by definition of a $n = 3$ polytrope, however $R$ is not the surface of the star. Here we will define the surface of the star $r_{0}$ as the radius from which an emitted photon has an optical depth of $\frac{2}{3}$\footnote{Using the Eddington gray approximation the effective temperature is equal to the temperature of the star at optical depth $\tau = \frac{2}{3}$.} . Here we assume Kramer's opacity law  for the gas at the surface parametrized by $\kappa = \kappa_{0} \rho T^{-7/2} $. Note that Kramer's opacity law assumes free-free and bound-free scattering as dominant where $\kappa_{0}$ is determined the dominant process. Here we ignore $H^{-}$ opacity and electron scattering opacity because the surface temperatures we will deal with are higher than $H^{-}$ binding energy and lower than the temperatures at which Compton scattering dominates. We will determine if these assumptions are justified in what follows. The scattering rate is then given by $\alpha_{K} = \kappa \rho$ and the optical depth at radius r is given by
\begin{align}
    \label{eq:tau}
    \tau(r) = \int_{r}^{R} \alpha_{K}(r) dr = \int_{r}^{R} \kappa_{0} \rho(r)^{2} T(r)^{-7/2}  dr = \int_{z}^{z_{3}} \kappa_{0} \frac{1}{A \xi^{7/2}} \rho_{c}^{5/6}  w(z)^{5/2} dz \,.
\end{align}
Let us assume $\kappa_{0} = f \times  0.38 \times \frac{\alpha^{3}_{D}}{m_{c}^{3/2}m^{2}_{X}}$ where $f = 1$ corresponds to free free opacity. Let $z_{0}$ be the $z$ at which the optical depth is $\frac{2}{3}$ which we define to be the surface of the star. We will assume $\delta z_{0} = z_{3}-z_{0} \ll 1$ \footnote{This is done to get an analytical approximation for surface and it will be justified in what follows.}. We can Taylor expand $w(z) = - w'(z)|_{z = z_{3}} \delta z_{0} = 0.042 \,\delta z_{0}$, then
\begin{align}
    \label{eq:taufin}
    \tau = 2.5 \times 10^{-5} \times f \times \left(\frac{M_{\rm pl}\alpha^{2}_{D}\sqrt{\mu \rho_{c}}}{\left(m_{c} m_{X}\right)^{3/2}}\right)\left(\frac{\sqrt{\beta}}{1-\beta}\right) \left(\delta z_{0}\right)^{7/2}  = \frac{2}{3}\,.
\end{align}
This gives us 
\begin{align}
    \label{eq:dz0}
    \delta z_{0} = 8.7 \times 10^{-3} \left(\frac{100}{f}\right)^{2/7} \left(\frac{m_{c}}{m_{e}}\right)^{3/7} \left(\frac{m_{X}}{m_{p}}\right)^{3/7}\left(\frac{\alpha}{\alpha_{D}}\right)^{6/7} \left(\frac{1-\beta}{\beta^{1/2}}\right)^{2/7} \left(\frac{10^{3} \rm{gm~cm^{-3}}}{\rho_{c}}\right)^{1/7}\,.
\end{align}
Note that the benchmark value $\left(\rho_{c} = 10^{3}\rm{gm~cm^{-3}}\right)$ we use above is the density around at which core collapse begins for SM parameters. Here we assume $f = 100$ (which is the case in SM) i.e. the Rosseland mean for the atomic opacity is assumed to be $10^{2}$ times larger than free-free opacity \footnote{$f = 100$ is a conservative estimate as larger $f$ implies lower temperature and hence a longer cooling time.}.
Then we get
\begin{multline}
    \label{eq:Ts}
    T_{s} = T(r_{0}) = \xi \rho^{1/3}_{c} w(z_{0})
    = 1.29 \times 10^{-8}\  \GeV \left(\frac{100}{f}\right)^{2/7} \left(\frac{m_{c}}{m_{e}}\right)^{3/7} \\
    \times \left(\frac{m_{X}} {m_{p}}\right)^{2/21}
    \left(\frac{\alpha}{\alpha_{D}}\right)^{6/7}
    \left(\frac{(1-\beta)^{13}}{\beta^{10}}\right)^{1/21}
    \left(\frac{\rho_{c}}{10^{3} \rm{gm~cm^{-3}}}\right)^{4/21}\,.
\end{multline}
For SM parameters for a $M_{*} = 60 \Msol$ we get $\beta = 0.75$. Using the benchmark density and $f = 100$, we get $T_{s} = 72,000 K$. This compares well to the value from a MESA simulation which gives $10^{5} K$.The small difference can be accounted for by the several approximations we made \footnote{The photosphere of a star is also effectively a $n=3.25$ polytrope, not a $n=3$ polytrope as we assumed. We also assumed the opacity was defined at a hard cutoff (Eq.~\ref{eq:taufin}) of the optical depth to which our final answer is sensitive.}. 
As shown in Fig.~\ref{fig:Tsurf100}, our analytical approximation of the surface density matches reasonably well in slope as well as magnitude when compared to the exact results from MESA. 

\subsection{Equation of State and Dynamical Instability}
\label{sec:eos}

The calculation of $\Gamma_1$ requires computation of the complete equation of state. We note that the adiabatic index is, by definition, a local quantity. This means that even if a particular region of the "star" is dynamically unstable, it is possible that the system as a whole is stable. Ideally, one should use a global quantity to describe the system, such as the pressure-weighted averaged adiabatic index (see e.g. \cite{renzo_predictions_2020})\footnote{For the derivation of the adiabatic index in the context of self-interacting dark matter look Ref~\cite{Feng:2021rst}.}

\begin{equation}
    \Big \langle \Gamma_1 \Big \rangle \equiv \frac{\int \Gamma_1 P dV}{\int P dV}.
\end{equation}
However, in this work, we consider the local adiabatic index, evaluated at the center of the star for simplicity. We consider a star experiences pair instability and collapse for any $M > M_{\rm crit}$, where $M_{\rm crit}$ is the lowest mass star whose adiabatic index in the core drops below $4/3$. 
\begin{figure}[!t]
\centering
\includegraphics[width=0.71\textwidth]{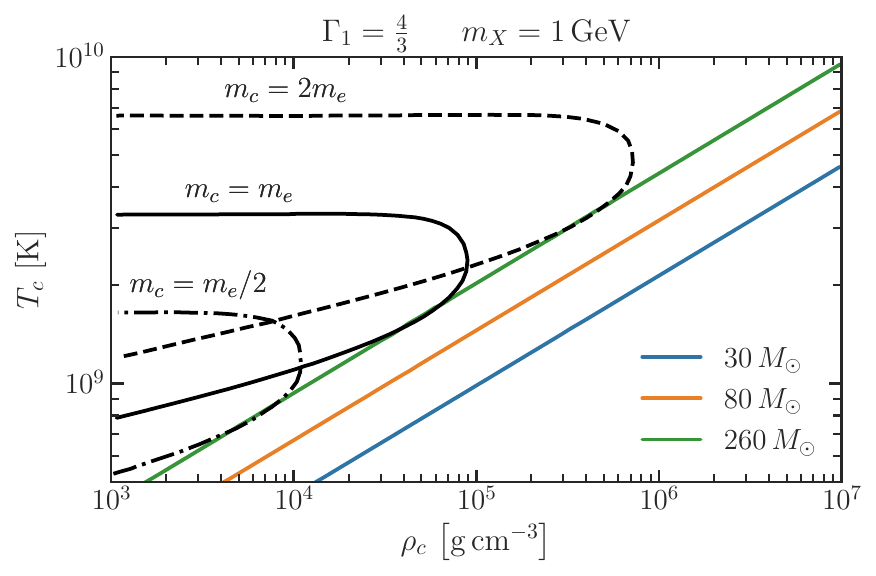}
    \caption{Dynamic instability regions and polytropic evolution lines for $m_X = 1\GeV$.}
    \label{fig:gamma1}
\end{figure}
The calculation of the ion and radiation contribution to the total pressure is straightforward for a given density and temperature as follows. The contribution to the total pressure and entropy from radiation is 
\begin{align}
    P_\mathrm{rad} &= \frac{a}{3}T^4 \\
    s_\mathrm{rad} &= \frac{4}{3} a \frac{T^3}{\rho}\,,
\end{align}
while the contribution to the total pressure and entropy from ions is 
\begin{align}
    P_\mathrm{ion} &= \Big\langle \frac{1}{A} \Big\rangle \frac{\rho}{m_X} k_B T \\
    s_\mathrm{ion} &= \Big\langle\frac{1}{A} \Big\rangle \frac{k_B}{m_X}\Big(\frac{5}{2} + \log(\frac{T^{3/2}}{\rho}) \Big)\,,
\end{align}
where $\Big\langle\frac{1}{A}\Big\rangle$ is the mean reciprocal atomic weight, which for ionized (dark) hydrogen is $\approx 1$. Note that the electron contribution will be dealt with separately to allow for relativistic and quantum effects. 
The contribution from dark electrons and positrons , $P_c$, requires integration over the density of states but is numerically tractable. We outline this procedure here for completeness, following \cite{Rakavy_1967, Croon:2020oga}.
We define the following quantities, 
\begin{equation}
     C = \frac{1}{\pi^2}\Big( \frac{m_c c}{\hbar}\Big)^3, \pad \beta = \frac{m_c c^2}{k_B T}, \pad \phi = \frac{\mu}{k_B T}\,,
\end{equation}
where $\mu$ is the chemical potential of the negatively charged charge carrier. The chemical potential is determined by imposing charge neutrality and solving for the roots of the excess charge carrier number density, satisfying
\begin{equation}
    n_c(\phi, \beta) = n_c^- - n_c^+ = \Big \langle \frac{Z}{A} \Big \rangle \frac{\rho}{m_X}.
\end{equation}
With these definitions, the number density, pressure, and entropy are given by 
\begin{align}
    n_c(\phi, \beta) &= m_c C F_2^+(\phi, \beta)\,,\\
    P_c(\phi, \beta) &= m_c c^2 C F_1(\phi, \beta)\,,\\
    s_c(\phi, \beta) &= \frac{k_B C \beta}{\rho}\Big(F_1(\phi, \beta) + F_3(\phi, \beta) - \frac{\phi}{\beta} F_2^-(\phi, \beta) \Big)\,,
\end{align}
where the modified incomplete Fermi-Dirac integrals that appear in the previous definitions are given by:
\begin{align}
    F_1(\phi, \beta) &= \int_{\varepsilon = \beta}^\infty \Gamma \Big( \frac{\varepsilon}{\beta}\Big) D^+(\varepsilon, \phi) \frac{d\varepsilon}{\beta}\,, \\
    F_2^+(\phi, \beta) &= \int_{\varepsilon = \beta}^\infty \Gamma' \Big( \frac{\varepsilon}{\beta}\Big) D^+(\varepsilon, \phi) \frac{d\varepsilon}{\beta}\,, \\
    F_2^-(\phi, \beta) &= \int_{\varepsilon = \beta}^\infty \Gamma' \Big( \frac{\varepsilon}{\beta}\Big) D^-(\varepsilon, \phi) \frac{d\varepsilon}{\beta}\,, \\
    F_3(\phi, \beta) &= \int_{\varepsilon = \beta}^\infty \varepsilon \Gamma' \Big( \frac{\varepsilon}{\beta}\Big) D^+(\varepsilon, \phi) \frac{d\varepsilon}{\beta^2}\,,
\end{align}
where $\Gamma(x) \equiv \frac{1}{3}(x^2 - 1)^{3/2}$ and $\varepsilon \equiv \frac{E}{k_B T}$. The adiabatic index (\ref{eq:gamma1}) can be written as
\begin{equation}
    \frac{\rho}{P} \Big( \frac{\partial P}{\partial \rho}\Big)_s = \frac{\rho}{P} \Big[ \Big(\frac{\partial P}{\partial \rho}\Big)_T + \Big(\frac{\partial P}{\partial T}\Big)_\rho \Big(\frac{\partial T}{\partial \rho}\Big)_s \Big]\,,
\end{equation}
which can then be calculated numerically term-by-term from the above expressions.
\begin{figure}[!t]
\centering
\includegraphics[width=0.70\textwidth]{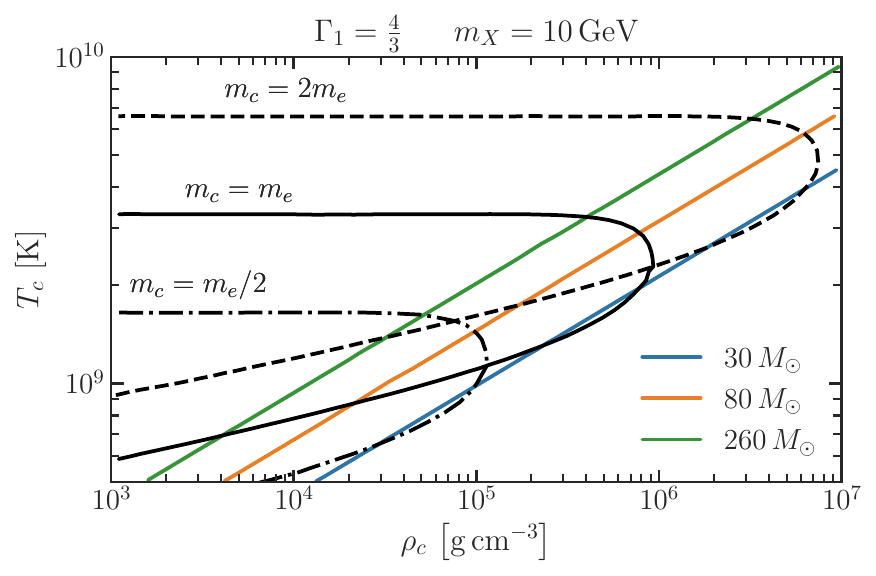}
    \caption{Dynamic instability regions and polytropic evolution lines for $m_X = 10 \GeV$.}
    \label{fig:gamma110mx}
\end{figure}
Under what conditions will a dark star enter the dynamic instability region and subsequently collapse? It turns out that, assuming an $n=3$ polytropic evolution, whether or not a dark star will enter the region of dynamic instability is independent of the charge carrier mass and dark fine-structure constant, $m_c, \alpha_{D}$ and only depends on the dark proton mass, $m_{X}$. Although the region where $\Gamma_1 < 4/3$ varies with $m_c$, a single line corresponding to $T \sim \rho^{1/3}$ is tangent to all regions.  In other words, even though the density and temperature at which a star would enter the $\Gamma_1 < 4/3$ region depends on $m_c$, the ultimate fate of the dark star is independent of the charge carrier mass, given sufficient time for the system to evolve. 
In contrast, the final fate of a dark star is strongly dependent on the dark proton mass, $m_X$. For larger $m_X$, lighter stars will pass through the dynamic instability region throughout their evolution, causing them to collapse. For $m_X = 10 \ \GeV$, a star with $M \gtrsim 30 \Msol$ will become dynamically unstable. Empirically, the critical mass scales as 
\begin{equation}
    M_\mathrm{crit} \approx 260 \Msol \Big( \frac{\text{GeV}}{m_X}\Big)\,.
\end{equation}
Here, we assume that all dark stars with sufficiently high masses will collapse. 

\bibliographystyle{JHEP}
\bibliography{references}

\providecommand{\href}[2]{#2}\begingroup\raggedright\begin{thebibliography}{10}

\bibitem{PhysRevLett.116.061102}
{\scshape LIGO Scientific Collaboration and Virgo Collaboration} collaboration,
  B.~P. Abbott, R.~Abbott, T.~D. Abbott, M.~R. Abernathy, F.~Acernese,
  K.~Ackley et~al., \emph{Observation of gravitational waves from a binary
  black hole merger},
  \href{https://doi.org/10.1103/PhysRevLett.116.061102}{\emph{Phys. Rev. Lett.}
  {\bfseries 116} (2016) 061102}.

\bibitem{LIGOScientific:2021psn}
{\scshape LIGO Scientific, VIRGO, KAGRA} collaboration, R.~Abbott et~al.,
  \emph{{The population of merging compact binaries inferred using
  gravitational waves through GWTC-3}},
  \href{https://arxiv.org/abs/2111.03634}{{\ttfamily 2111.03634}}.

\bibitem{Woosley_2021}
S.~E. Woosley and A.~Heger, \emph{The pair-instability mass gap for black
  holes}, \href{https://doi.org/10.3847/2041-8213/abf2c4}{\emph{The
  Astrophysical Journal Letters} {\bfseries 912} (2021) L31}.

\bibitem{Carr:2009jm}
B.~J. Carr, K.~Kohri, Y.~Sendouda and J.~Yokoyama, \emph{{New cosmological
  constraints on primordial black holes}},
  \href{https://doi.org/10.1103/PhysRevD.81.104019}{\emph{Phys. Rev. D}
  {\bfseries 81} (2010) 104019}
  [\href{https://arxiv.org/abs/0912.5297}{{\ttfamily 0912.5297}}].

\bibitem{Ashoorioon:2019xqc}
A.~Ashoorioon, A.~Rostami and J.~T. Firouzjaee, \emph{{EFT compatible PBHs:
  effective spawning of the seeds for primordial black holes during
  inflation}}, \href{https://doi.org/10.1007/JHEP07(2021)087}{\emph{JHEP}
  {\bfseries 07} (2021) 087}
  [\href{https://arxiv.org/abs/1912.13326}{{\ttfamily 1912.13326}}].

\bibitem{Ashoorioon:2022raz}
A.~Ashoorioon, K.~Rezazadeh and A.~Rostami, \emph{{NANOGrav Signal from the End
  of Inflation and the LIGO Mass and Heavier Primordial Black Holes}},
  \href{https://arxiv.org/abs/2202.01131}{{\ttfamily 2202.01131}}.

\bibitem{Shandera:2018xkn}
S.~Shandera, D.~Jeong and H.~S.~G. Gebhardt, \emph{{Gravitational Waves from
  Binary Mergers of Subsolar Mass Dark Black Holes}},
  \href{https://doi.org/10.1103/PhysRevLett.120.241102}{\emph{Phys. Rev. Lett.}
  {\bfseries 120} (2018) 241102}
  [\href{https://arxiv.org/abs/1802.08206}{{\ttfamily 1802.08206}}].

\bibitem{Rakavy_1967}
G.~{Rakavy} and G.~{Shaviv}, \emph{{Instabilities in Highly Evolved Stellar
  Models}}, \href{https://doi.org/10.1086/149204}{\emph{apj} {\bfseries 148}
  (1967) 803}.

\bibitem{PhysRevLett.18.379}
Z.~Barkat, G.~Rakavy and N.~Sack, \emph{Dynamics of supernova explosion
  resulting from pair formation},
  \href{https://doi.org/10.1103/PhysRevLett.18.379}{\emph{Phys. Rev. Lett.}
  {\bfseries 18} (1967) 379}.

\bibitem{Woosley_2017}
S.~E. Woosley, \emph{Pulsational pair-instability supernovae},
  \href{https://doi.org/10.3847/1538-4357/836/2/244}{\emph{The Astrophysical
  Journal} {\bfseries 836} (2017) 244}.

\bibitem{Heger_2002}
A.~Heger and S.~E. Woosley, \emph{The nucleosynthetic signature of population
  {III}}, \href{https://doi.org/10.1086/338487}{\emph{The Astrophysical
  Journal} {\bfseries 567} (2002) 532}.

\bibitem{LIGOScientific:2014pky}
{\scshape LIGO Scientific} collaboration, J.~Aasi et~al., \emph{{Advanced
  LIGO}}, \href{https://doi.org/10.1088/0264-9381/32/7/074001}{\emph{Class.
  Quant. Grav.} {\bfseries 32} (2015) 074001}
  [\href{https://arxiv.org/abs/1411.4547}{{\ttfamily 1411.4547}}].

\bibitem{VIRGO:2014yos}
{\scshape VIRGO} collaboration, F.~Acernese et~al., \emph{{Advanced Virgo: a
  second-generation interferometric gravitational wave detector}},
  \href{https://doi.org/10.1088/0264-9381/32/2/024001}{\emph{Class. Quant.
  Grav.} {\bfseries 32} (2015) 024001}
  [\href{https://arxiv.org/abs/1408.3978}{{\ttfamily 1408.3978}}].

\bibitem{LIGOScientific:2021usb}
{\scshape LIGO Scientific, VIRGO} collaboration, R.~Abbott et~al.,
  \emph{{GWTC-2.1: Deep Extended Catalog of Compact Binary Coalescences
  Observed by LIGO and Virgo During the First Half of the Third Observing
  Run}},  \href{https://arxiv.org/abs/2108.01045}{{\ttfamily 2108.01045}}.

\bibitem{LIGOScientific:2021djp}
{\scshape LIGO Scientific, VIRGO, KAGRA} collaboration, R.~Abbott et~al.,
  \emph{{GWTC-3: Compact Binary Coalescences Observed by LIGO and Virgo During
  the Second Part of the Third Observing Run}},
  \href{https://arxiv.org/abs/2111.03606}{{\ttfamily 2111.03606}}.

\bibitem{Gerosa:2017kvu}
D.~Gerosa and E.~Berti, \emph{{Are merging black holes born from stellar
  collapse or previous mergers?}},
  \href{https://doi.org/10.1103/PhysRevD.95.124046}{\emph{Phys. Rev. D}
  {\bfseries 95} (2017) 124046}
  [\href{https://arxiv.org/abs/1703.06223}{{\ttfamily 1703.06223}}].

\bibitem{Fishbach:2017dwv}
M.~Fishbach, D.~E. Holz and B.~Farr, \emph{{Are LIGO's Black Holes Made From
  Smaller Black Holes?}},
  \href{https://doi.org/10.3847/2041-8213/aa7045}{\emph{Astrophys. J. Lett.}
  {\bfseries 840} (2017) L24}
  [\href{https://arxiv.org/abs/1703.06869}{{\ttfamily 1703.06869}}].

\bibitem{Rodriguez:2019huv}
C.~L. Rodriguez, M.~Zevin, P.~Amaro-Seoane, S.~Chatterjee, K.~Kremer, F.~A.
  Rasio et~al., \emph{{Black holes: The next generation\textemdash{}repeated
  mergers in dense star clusters and their gravitational-wave properties}},
  \href{https://doi.org/10.1103/PhysRevD.100.043027}{\emph{Phys. Rev. D}
  {\bfseries 100} (2019) 043027}
  [\href{https://arxiv.org/abs/1906.10260}{{\ttfamily 1906.10260}}].

\bibitem{LIGOScientific:2020ufj}
{\scshape LIGO Scientific, Virgo} collaboration, R.~Abbott et~al.,
  \emph{{Properties and Astrophysical Implications of the 150 M$_\odot$ Binary
  Black Hole Merger GW190521}},
  \href{https://doi.org/10.3847/2041-8213/aba493}{\emph{Astrophys. J. Lett.}
  {\bfseries 900} (2020) L13}
  [\href{https://arxiv.org/abs/2009.01190}{{\ttfamily 2009.01190}}].

\bibitem{Roupas:2018cvb}
Z.~Roupas and D.~Kazanas, \emph{{Binary black hole growth by gas accretion in
  stellar clusters}},
  \href{https://doi.org/10.1051/0004-6361/201834609}{\emph{Astron. Astrophys.}
  {\bfseries 621} (2019) L1}
  [\href{https://arxiv.org/abs/1809.04126}{{\ttfamily 1809.04126}}].

\bibitem{Roupas:2019dgx}
Z.~Roupas and D.~Kazanas, \emph{{Generation of massive stellar black holes by
  rapid gas accretion in primordial dense clusters}},
  \href{https://doi.org/10.1051/0004-6361/201937002}{\emph{Astron. Astrophys.}
  {\bfseries 632} (2019) L8}
  [\href{https://arxiv.org/abs/1911.03915}{{\ttfamily 1911.03915}}].

\bibitem{vanSon:2020zbk}
L.~A.~C. van Son, S.~E. de~Mink, F.~S. Broekgaarden, M.~Renzo, S.~Justham,
  E.~Laplace et~al., \emph{{Polluting the pair-instability mass gap for binary
  black holes through super-Eddington accretion in isolated binaries}},
  \href{https://doi.org/10.3847/1538-4357/ab9809}{\emph{Astrophys. J.}
  {\bfseries 897} (2020) 100}
  [\href{https://arxiv.org/abs/2004.05187}{{\ttfamily 2004.05187}}].

\bibitem{Spera:2018wnw}
M.~Spera, M.~Mapelli, N.~Giacobbo, A.~A. Trani, A.~Bressan and G.~Costa,
  \emph{{Merging black hole binaries with the SEVN code}},
  \href{https://arxiv.org/abs/1809.04605}{{\ttfamily 1809.04605}}.

\bibitem{DiCarlo:2019pmf}
U.~N. Di~Carlo, N.~Giacobbo, M.~Mapelli, M.~Pasquato, M.~Spera, L.~Wang et~al.,
  \emph{{Merging black holes in young star clusters}},
  \href{https://doi.org/10.1093/mnras/stz1453}{\emph{Mon. Not. Roy. Astron.
  Soc.} {\bfseries 487} (2019) 2947}
  [\href{https://arxiv.org/abs/1901.00863}{{\ttfamily 1901.00863}}].

\bibitem{Croon:2020ehi}
D.~Croon, S.~D. McDermott and J.~Sakstein, \emph{{Missing in axion: Where are
  XENON1T\textquoteright{}s big black holes?}},
  \href{https://doi.org/10.1016/j.dark.2021.100801}{\emph{Phys. Dark Univ.}
  {\bfseries 32} (2021) 100801}
  [\href{https://arxiv.org/abs/2007.00650}{{\ttfamily 2007.00650}}].

\bibitem{Croon:2020oga}
D.~Croon, S.~D. McDermott and J.~Sakstein, \emph{{New physics and the black
  hole mass gap}},
  \href{https://doi.org/10.1103/PhysRevD.102.115024}{\emph{Phys. Rev. D}
  {\bfseries 102} (2020) 115024}
  [\href{https://arxiv.org/abs/2007.07889}{{\ttfamily 2007.07889}}].

\bibitem{Sakstein:2020axg}
J.~Sakstein, D.~Croon, S.~D. McDermott, M.~C. Straight and E.~J. Baxter,
  \emph{{Beyond the Standard Model Explanations of GW190521}},
  \href{https://doi.org/10.1103/PhysRevLett.125.261105}{\emph{Phys. Rev. Lett.}
  {\bfseries 125} (2020) 261105}
  [\href{https://arxiv.org/abs/2009.01213}{{\ttfamily 2009.01213}}].

\bibitem{Ziegler:2020klg}
J.~Ziegler and K.~Freese, \emph{{Filling the black hole mass gap: Avoiding pair
  instability in massive stars through addition of nonnuclear energy}},
  \href{https://doi.org/10.1103/PhysRevD.104.043015}{\emph{Phys. Rev. D}
  {\bfseries 104} (2021) 043015}
  [\href{https://arxiv.org/abs/2010.00254}{{\ttfamily 2010.00254}}].

\bibitem{Chacko:2005pe}
Z.~Chacko, H.-S. Goh and R.~Harnik, \emph{{The Twin Higgs: Natural electroweak
  breaking from mirror symmetry}},
  \href{https://doi.org/10.1103/PhysRevLett.96.231802}{\emph{Phys. Rev. Lett.}
  {\bfseries 96} (2006) 231802}
  [\href{https://arxiv.org/abs/hep-ph/0506256}{{\ttfamily hep-ph/0506256}}].

\bibitem{Arkani-Hamed:2016rle}
N.~Arkani-Hamed, T.~Cohen, R.~T. D'Agnolo, A.~Hook, H.~D. Kim and D.~Pinner,
  \emph{{Solving the Hierarchy Problem at Reheating with a Large Number of
  Degrees of Freedom}},
  \href{https://doi.org/10.1103/PhysRevLett.117.251801}{\emph{Phys. Rev. Lett.}
  {\bfseries 117} (2016) 251801}
  [\href{https://arxiv.org/abs/1607.06821}{{\ttfamily 1607.06821}}].

\bibitem{Goldberg:1986nk}
H.~Goldberg and L.~J. Hall, \emph{{A New Candidate for Dark Matter}},
  \href{https://doi.org/10.1016/0370-2693(86)90731-8}{\emph{Phys. Lett. B}
  {\bfseries 174} (1986) 151}.

\bibitem{Kaplan:2009de}
D.~E. Kaplan, G.~Z. Krnjaic, K.~R. Rehermann and C.~M. Wells, \emph{{Atomic
  Dark Matter}},
  \href{https://doi.org/10.1088/1475-7516/2010/05/021}{\emph{JCAP} {\bfseries
  05} (2010) 021} [\href{https://arxiv.org/abs/0909.0753}{{\ttfamily
  0909.0753}}].

\bibitem{Kaplan:2011yj}
D.~E. Kaplan, G.~Z. Krnjaic, K.~R. Rehermann and C.~M. Wells, \emph{{Dark
  Atoms: Asymmetry and Direct Detection}},
  \href{https://doi.org/10.1088/1475-7516/2011/10/011}{\emph{JCAP} {\bfseries
  10} (2011) 011} [\href{https://arxiv.org/abs/1105.2073}{{\ttfamily
  1105.2073}}].

\bibitem{Cyr-Racine:2012tfp}
F.-Y. Cyr-Racine and K.~Sigurdson, \emph{{Cosmology of atomic dark matter}},
  \href{https://doi.org/10.1103/PhysRevD.87.103515}{\emph{Phys. Rev. D}
  {\bfseries 87} (2013) 103515}
  [\href{https://arxiv.org/abs/1209.5752}{{\ttfamily 1209.5752}}].

\bibitem{Cline:2012is}
J.~M. Cline, Z.~Liu and W.~Xue, \emph{{Millicharged Atomic Dark Matter}},
  \href{https://doi.org/10.1103/PhysRevD.85.101302}{\emph{Phys. Rev. D}
  {\bfseries 85} (2012) 101302}
  [\href{https://arxiv.org/abs/1201.4858}{{\ttfamily 1201.4858}}].

\bibitem{Cline:2013pca}
J.~M. Cline, Z.~Liu, G.~Moore and W.~Xue, \emph{{Scattering properties of dark
  atoms and molecules}},
  \href{https://doi.org/10.1103/PhysRevD.89.043514}{\emph{Phys. Rev. D}
  {\bfseries 89} (2014) 043514}
  [\href{https://arxiv.org/abs/1311.6468}{{\ttfamily 1311.6468}}].

\bibitem{Fan:2013yva}
J.~Fan, A.~Katz, L.~Randall and M.~Reece, \emph{{Double-Disk Dark Matter}},
  \href{https://doi.org/10.1016/j.dark.2013.07.001}{\emph{Phys. Dark Univ.}
  {\bfseries 2} (2013) 139} [\href{https://arxiv.org/abs/1303.1521}{{\ttfamily
  1303.1521}}].

\bibitem{Fan:2013tia}
J.~Fan, A.~Katz, L.~Randall and M.~Reece, \emph{{Dark-Disk Universe}},
  \href{https://doi.org/10.1103/PhysRevLett.110.211302}{\emph{Phys. Rev. Lett.}
  {\bfseries 110} (2013) 211302}
  [\href{https://arxiv.org/abs/1303.3271}{{\ttfamily 1303.3271}}].

\bibitem{Fan:2013bea}
J.~Fan, A.~Katz and J.~Shelton, \emph{{Direct and indirect detection of
  dissipative dark matter}},
  \href{https://doi.org/10.1088/1475-7516/2014/06/059}{\emph{JCAP} {\bfseries
  06} (2014) 059} [\href{https://arxiv.org/abs/1312.1336}{{\ttfamily
  1312.1336}}].

\bibitem{Cyr-Racine:2013fsa}
F.-Y. Cyr-Racine, R.~de~Putter, A.~Raccanelli and K.~Sigurdson,
  \emph{{Constraints on Large-Scale Dark Acoustic Oscillations from
  Cosmology}}, \href{https://doi.org/10.1103/PhysRevD.89.063517}{\emph{Phys.
  Rev. D} {\bfseries 89} (2014) 063517}
  [\href{https://arxiv.org/abs/1310.3278}{{\ttfamily 1310.3278}}].

\bibitem{2014}
L.~Randall and M.~Reece, \emph{Dark matter as a trigger for periodic comet
  impacts},
  \href{https://doi.org/10.1103/physrevlett.112.161301}{\emph{Physical Review
  Letters} {\bfseries 112} (2014) }.

\bibitem{Foot:2014osa}
R.~Foot and S.~Vagnozzi, \emph{{Diurnal modulation signal from dissipative
  hidden sector dark matter}},
  \href{https://doi.org/10.1016/j.physletb.2015.06.063}{\emph{Phys. Lett. B}
  {\bfseries 748} (2015) 61} [\href{https://arxiv.org/abs/1412.0762}{{\ttfamily
  1412.0762}}].

\bibitem{Foot:2014uba}
R.~Foot and S.~Vagnozzi, \emph{{Dissipative hidden sector dark matter}},
  \href{https://doi.org/10.1103/PhysRevD.91.023512}{\emph{Phys. Rev. D}
  {\bfseries 91} (2015) 023512}
  [\href{https://arxiv.org/abs/1409.7174}{{\ttfamily 1409.7174}}].

\bibitem{Foot:2016wvj}
R.~Foot and S.~Vagnozzi, \emph{{Solving the small-scale structure puzzles with
  dissipative dark matter}},
  \href{https://doi.org/10.1088/1475-7516/2016/07/013}{\emph{JCAP} {\bfseries
  07} (2016) 013} [\href{https://arxiv.org/abs/1602.02467}{{\ttfamily
  1602.02467}}].

\bibitem{Rosenberg:2017qia}
E.~Rosenberg and J.~Fan, \emph{{Cooling in a Dissipative Dark Sector}},
  \href{https://doi.org/10.1103/PhysRevD.96.123001}{\emph{Phys. Rev. D}
  {\bfseries 96} (2017) 123001}
  [\href{https://arxiv.org/abs/1705.10341}{{\ttfamily 1705.10341}}].

\bibitem{Ghalsasi:2017jna}
A.~Ghalsasi and M.~McQuinn, \emph{{Exploring the astrophysics of dark atoms}},
  \href{https://doi.org/10.1103/PhysRevD.97.123018}{\emph{Phys. Rev. D}
  {\bfseries 97} (2018) 123018}
  [\href{https://arxiv.org/abs/1712.04779}{{\ttfamily 1712.04779}}].

\bibitem{Chang:2018bgx}
J.~H. Chang, D.~Egana-Ugrinovic, R.~Essig and C.~Kouvaris, \emph{{Structure
  Formation and Exotic Compact Objects in a Dissipative Dark Sector}},
  \href{https://doi.org/10.1088/1475-7516/2019/03/036}{\emph{JCAP} {\bfseries
  03} (2019) 036} [\href{https://arxiv.org/abs/1812.07000}{{\ttfamily
  1812.07000}}].

\bibitem{Gresham:2018anj}
M.~I. Gresham, H.~K. Lou and K.~M. Zurek, \emph{{Astrophysical Signatures of
  Asymmetric Dark Matter Bound States}},
  \href{https://doi.org/10.1103/PhysRevD.98.096001}{\emph{Phys. Rev. D}
  {\bfseries 98} (2018) 096001}
  [\href{https://arxiv.org/abs/1805.04512}{{\ttfamily 1805.04512}}].

\bibitem{Essig:2018pzq}
R.~Essig, S.~D. Mcdermott, H.-B. Yu and Y.-M. Zhong, \emph{{Constraining
  Dissipative Dark Matter Self-Interactions}},
  \href{https://doi.org/10.1103/PhysRevLett.123.121102}{\emph{Phys. Rev. Lett.}
  {\bfseries 123} (2019) 121102}
  [\href{https://arxiv.org/abs/1809.01144}{{\ttfamily 1809.01144}}].

\bibitem{Alvarez:2019nwt}
G.~Alvarez and H.-B. Yu, \emph{{Astrophysical probes of inelastic dark matter
  with a light mediator}},
  \href{https://doi.org/10.1103/PhysRevD.101.043002}{\emph{Phys. Rev. D}
  {\bfseries 101} (2020) 043002}
  [\href{https://arxiv.org/abs/1911.11114}{{\ttfamily 1911.11114}}].

\bibitem{Roux:2020wkp}
J.-S. Roux and J.~M. Cline, \emph{{Constraining galactic structures of mirror
  dark matter}}, \href{https://doi.org/10.1103/PhysRevD.102.063518}{\emph{Phys.
  Rev. D} {\bfseries 102} (2020) 063518}
  [\href{https://arxiv.org/abs/2001.11504}{{\ttfamily 2001.11504}}].

\bibitem{Cyr-Racine:2021oal}
F.-Y. Cyr-Racine, F.~Ge and L.~Knox, \emph{{Symmetry of Cosmological
  Observables, a Mirror World Dark Sector, and the Hubble Constant}},
  \href{https://doi.org/10.1103/PhysRevLett.128.201301}{\emph{Phys. Rev. Lett.}
  {\bfseries 128} (2022) 201301}
  [\href{https://arxiv.org/abs/2107.13000}{{\ttfamily 2107.13000}}].

\bibitem{Cline:2021itd}
J.~M. Cline, \emph{{Dark atoms and composite dark matter}},
  \href{https://arxiv.org/abs/2108.10314}{{\ttfamily 2108.10314}}.

\bibitem{Chacko:2021vin}
Z.~Chacko, D.~Curtin, M.~Geller and Y.~Tsai, \emph{{Direct detection of mirror
  matter in Twin Higgs models}},
  \href{https://doi.org/10.1007/JHEP11(2021)198}{\emph{JHEP} {\bfseries 11}
  (2021) 198} [\href{https://arxiv.org/abs/2104.02074}{{\ttfamily
  2104.02074}}].

\bibitem{Ryan:2021dis}
M.~Ryan, J.~Gurian, S.~Shandera and D.~Jeong, \emph{{Molecular Chemistry for
  Dark Matter}},  \href{https://arxiv.org/abs/2106.13245}{{\ttfamily
  2106.13245}}.

\bibitem{Gurian:2021qhk}
J.~Gurian, D.~Jeong, M.~Ryan and S.~Shandera, \emph{{Molecular Chemistry for
  Dark Matter II: Recombination, Molecule Formation, and Halo Mass Function in
  Atomic Dark Matter}},  \href{https://arxiv.org/abs/2110.11964}{{\ttfamily
  2110.11964}}.

\bibitem{Ryan:2021tgw}
M.~Ryan, S.~Shandera, J.~Gurian and D.~Jeong, \emph{{Molecular Chemistry for
  Dark Matter III: DarkKROME}},
  \href{https://arxiv.org/abs/2110.11971}{{\ttfamily 2110.11971}}.

\bibitem{Howe:2021neq}
A.~Howe, J.~Setford, D.~Curtin and C.~D. Matzner, \emph{{How to search for
  mirror stars with Gaia}},
  \href{https://doi.org/10.1007/JHEP07(2022)059}{\emph{JHEP} {\bfseries 07}
  (2022) 059} [\href{https://arxiv.org/abs/2112.05766}{{\ttfamily
  2112.05766}}].

\bibitem{Blinov:2021mdk}
N.~Blinov, G.~Krnjaic and S.~W. Li, \emph{{Realistic model of dark atoms to
  resolve the Hubble tension}},
  \href{https://doi.org/10.1103/PhysRevD.105.095005}{\emph{Phys. Rev. D}
  {\bfseries 105} (2022) 095005}
  [\href{https://arxiv.org/abs/2108.11386}{{\ttfamily 2108.11386}}].

\bibitem{Bansal:2021dfh}
S.~Bansal, J.~H. Kim, C.~Kolda, M.~Low and Y.~Tsai, \emph{{Mirror twin Higgs
  cosmology: constraints and a possible resolution to the H$_{0}$ and S$_{8}$
  tensions}}, \href{https://doi.org/10.1007/JHEP05(2022)050}{\emph{JHEP}
  {\bfseries 05} (2022) 050}
  [\href{https://arxiv.org/abs/2110.04317}{{\ttfamily 2110.04317}}].

\bibitem{Cruz:2022otv}
A.~Cruz and M.~McQuinn, \emph{{Astrophysical Plasma Instabilities induced by
  Long-Range Interacting Dark Matter}},
  \href{https://arxiv.org/abs/2202.12464}{{\ttfamily 2202.12464}}.

\bibitem{Peled:2022byr}
G.~Peled and T.~Volansky, \emph{{Constraining Dark Matter Inside Stars Using
  Spectroscopic Binaries and a Modified Mass-Luminosity Relation}},
  \href{https://arxiv.org/abs/2203.09522}{{\ttfamily 2203.09522}}.

\bibitem{Roy:2023zar}
S.~Roy, X.~Shen, M.~Lisanti, D.~Curtin, N.~Murray and P.~F. Hopkins,
  \emph{{Simulating Atomic Dark Matter in Milky Way Analogs}},
  \href{https://doi.org/10.3847/2041-8213/ace2c8}{\emph{Astrophys. J. Lett.}
  {\bfseries 954} (2023) L40}
  [\href{https://arxiv.org/abs/2304.09878}{{\ttfamily 2304.09878}}].

\bibitem{Gemmell:2023trd}
C.~Gemmell, S.~Roy, X.~Shen, D.~Curtin, M.~Lisanti, N.~Murray et~al.,
  \emph{{Dissipative Dark Substructure: The Consequences of Atomic Dark Matter
  on Milky Way Analog Subhalos}},
  \href{https://arxiv.org/abs/2311.02148}{{\ttfamily 2311.02148}}.

\bibitem{Kouvaris:2015rea}
C.~Kouvaris and N.~G. Nielsen, \emph{{Asymmetric Dark Matter Stars}},
  \href{https://doi.org/10.1103/PhysRevD.92.063526}{\emph{Phys. Rev. D}
  {\bfseries 92} (2015) 063526}
  [\href{https://arxiv.org/abs/1507.00959}{{\ttfamily 1507.00959}}].

\bibitem{Giudice:2016zpa}
G.~F. Giudice, M.~McCullough and A.~Urbano, \emph{{Hunting for Dark Particles
  with Gravitational Waves}},
  \href{https://doi.org/10.1088/1475-7516/2016/10/001}{\emph{JCAP} {\bfseries
  10} (2016) 001} [\href{https://arxiv.org/abs/1605.01209}{{\ttfamily
  1605.01209}}].

\bibitem{Curtin:2019lhm}
D.~Curtin and J.~Setford, \emph{{How To Discover Mirror Stars}},
  \href{https://doi.org/10.1016/j.physletb.2020.135391}{\emph{Phys. Lett. B}
  {\bfseries 804} (2020) 135391}
  [\href{https://arxiv.org/abs/1909.04071}{{\ttfamily 1909.04071}}].

\bibitem{Hippert:2021fch}
M.~Hippert, J.~Setford, H.~Tan, D.~Curtin, J.~Noronha-Hostler and N.~Yunes,
  \emph{{Mirror neutron stars}},
  \href{https://doi.org/10.1103/PhysRevD.106.035025}{\emph{Phys. Rev. D}
  {\bfseries 106} (2022) 035025}
  [\href{https://arxiv.org/abs/2103.01965}{{\ttfamily 2103.01965}}].

\bibitem{Gross:2021qgx}
C.~Gross, G.~Landini, A.~Strumia and D.~Teresi, \emph{{Dark Matter as dark
  dwarfs and other macroscopic objects: multiverse relics?}},
  \href{https://doi.org/10.1007/JHEP09(2021)033}{\emph{JHEP} {\bfseries 09}
  (2021) 033} [\href{https://arxiv.org/abs/2105.02840}{{\ttfamily
  2105.02840}}].

\bibitem{Ryan:2022hku}
M.~Ryan and D.~Radice, \emph{{Exotic compact objects: The dark white dwarf}},
  \href{https://doi.org/10.1103/PhysRevD.105.115034}{\emph{Phys. Rev. D}
  {\bfseries 105} (2022) 115034}
  [\href{https://arxiv.org/abs/2201.05626}{{\ttfamily 2201.05626}}].

\bibitem{Gurian:2022nbx}
J.~Gurian, M.~Ryan, S.~Schon, D.~Jeong and S.~Shandera, \emph{{A Lower Bound on
  the Mass of Compact Objects from Dissipative Dark Matter}},
  \href{https://doi.org/10.3847/2041-8213/ac997c}{\emph{Astrophys. J. Lett.}
  {\bfseries 939} (2022) L12}
  [\href{https://arxiv.org/abs/2209.00064}{{\ttfamily 2209.00064}}].

\bibitem{Armstrong:2023cis}
I.~Armstrong, B.~Gurbuz, D.~Curtin and C.~Matzner, \emph{{Electromagnetic
  Signatures of Mirror Stars}},
  \href{https://arxiv.org/abs/2311.18086}{{\ttfamily 2311.18086}}.

\bibitem{bromm_formation_2002}
V.~Bromm, P.~S. Coppi and R.~B. Larson, \emph{The {Formation} of the {First}
  {Stars}. {I}. {The} {Primordial} {Star} {Forming} {Cloud}},
  \href{https://doi.org/10.1086/323947}{\emph{The Astrophysical Journal}
  {\bfseries 564} (2002) 23}.

\bibitem{Schutz:2017tfp}
K.~Schutz, T.~Lin, B.~R. Safdi and C.-L. Wu, \emph{{Constraining a Thin Dark
  Matter Disk with Gaia}},
  \href{https://doi.org/10.1103/PhysRevLett.121.081101}{\emph{Phys. Rev. Lett.}
  {\bfseries 121} (2018) 081101}
  [\href{https://arxiv.org/abs/1711.03103}{{\ttfamily 1711.03103}}].

\bibitem{Brandt:2016aco}
T.~D. Brandt, \emph{{Constraints on MACHO Dark Matter from Compact Stellar
  Systems in Ultra-Faint Dwarf Galaxies}},
  \href{https://doi.org/10.3847/2041-8205/824/2/L31}{\emph{Astrophys. J. Lett.}
  {\bfseries 824} (2016) L31}
  [\href{https://arxiv.org/abs/1605.03665}{{\ttfamily 1605.03665}}].

\bibitem{MACHO:2000qbb}
{\scshape MACHO} collaboration, C.~Alcock et~al., \emph{{The MACHO project:
  Microlensing results from 5.7 years of LMC observations}},
  \href{https://doi.org/10.1086/309512}{\emph{Astrophys. J.} {\bfseries 542}
  (2000) 281} [\href{https://arxiv.org/abs/astro-ph/0001272}{{\ttfamily
  astro-ph/0001272}}].

\bibitem{2015ApJS..216...12W}
{\L}.~{Wyrzykowski}, A.~E. {Rynkiewicz}, J.~{Skowron}, S.~{Koz{\l}owski},
  A.~{Udalski}, M.~K. {Szyma{\'n}ski} et~al., \emph{{OGLE-III Microlensing
  Events and the Structure of the Galactic Bulge}},
  \href{https://doi.org/10.1088/0067-0049/216/1/12}{\emph{\apjs} {\bfseries
  216} (2015) 12} [\href{https://arxiv.org/abs/1405.3134}{{\ttfamily
  1405.3134}}].

\bibitem{1974ApJ...187..425P}
W.~H. {Press} and P.~{Schechter}, \emph{{Formation of Galaxies and Clusters of
  Galaxies by Self-Similar Gravitational Condensation}},
  \href{https://doi.org/10.1086/152650}{\emph{\apj} {\bfseries 187} (1974)
  425}.

\bibitem{1991ApJ...379..440B}
J.~R. {Bond}, S.~{Cole}, G.~{Efstathiou} and N.~{Kaiser}, \emph{{Excursion Set
  Mass Functions for Hierarchical Gaussian Fluctuations}},
  \href{https://doi.org/10.1086/170520}{\emph{\apj} {\bfseries 379} (1991)
  440}.

\bibitem{Lacey:1993iv}
C.~G. Lacey and S.~Cole, \emph{{Merger rates in hierarchical models of galaxy
  formation}}, {\emph{Mon. Not. Roy. Astron. Soc.} {\bfseries 262} (1993) 627}.

\bibitem{renzo_predictions_2020}
M.~Renzo, R.~Farmer, S.~Justham, Y.~Götberg, S.~E. de~Mink, E.~Zapartas
  et~al., \emph{Predictions for the hydrogen-free ejecta of pulsational
  pair-instability supernovae},
  \href{https://doi.org/10.1051/0004-6361/202037710}{\emph{Astronomy \&
  Astrophysics} {\bfseries 640} (2020) A56}.

\bibitem{Feng:2021rst}
W.-X. Feng, H.-B. Yu and Y.-M. Zhong, \emph{{Dynamical instability of collapsed
  dark matter halos}},
  \href{https://doi.org/10.1088/1475-7516/2022/05/036}{\emph{JCAP} {\bfseries
  05} (2022) 036} [\href{https://arxiv.org/abs/2108.11967}{{\ttfamily
  2108.11967}}].

\end{thebibliography}\endgroup
\end{document}